# Transitions toward Quantum Chaos: with Supersymmetry from Poisson to Gauss

Thomas Guhr
Center for Chaos and Turbulence Studies
Niels Bohr Institute
Blegdamsvej 17, 2100 Copenhagen Ø, Denmark
and
Max Planck Institut für Kernphysik
Saupfercheckweg 1, 69117 Heidelberg, Germany

September 27, 1995

**Abstract**

The transition from arbitrary to chaotic fluctuation properties in quantum systems is studied in a random matrix model. It is assumed that the Hamiltonian can be written as the sum of an arbitrary and a chaos producing part. The Gaussian ensembles are used to model the chaotic part. A closed integral representation for all correlations in the case of broken time reversal invariance is derived by employing supersymmetry and the graded eigenvalue method. In particular, the two level correlation function is expressed as a double integral. For a correlation index, exact diffusion equations are derived for all three universality classes which describe the transition to the chaotic regime from arbitrary initial conditions. As an application, the transition from Poisson regularity to chaos is discussed. The two level correlation function becomes a double integral for all values of the transition parameter.

# 1  Introduction

In classical and quantum mechanics, the interaction of a system is very often a sum of different contributions whose individual properties are well understood. The Hydrogen atom and the electron in a magnetic field, for example, are two of the best studied systems in quantum physics. The Hydrogen atom in a strong magnetic field, however, the combination of these two systems, could only fairly recently be investigated in full detail [1]. The total interaction is a sum of three competing terms, the Coulomb, the paramagnetic and the diamagnetic one. The influence of the different contributions can most instructively be seen when analyzing observables that are sensitive to regular or chaotic behavior of the system. We use the designation "regular" or "chaotic" for a quantum system whose classical analogue is integrable and regular or non-integrable and chaotic, respectively. Observables of particular interest for those studies are the nearest neighbor spacing distribution and the spectral rigidity [2]. In the case of the Hydrogen atom in a strong magnetic field, these two observables approach the Wigner-Dyson predictions [2, 3, 4, 5] gradually as the magnetic field strength is increased, and thus the system makes a transition from a regular to a chaotic one [1]. According to this observation, the Hamiltonian of a statistical model describing a regular system in an external field which induces chaotic motion can be written as the sum of a regular and a chaotic part. The theory of random matrices [3] provides the Gaussian Ensembles as well understood models for the chaotic part. Due to the general symmetry constraints, a time reversal invariant system with conserved or broken rotation invariance is modeled by the Gaussian Orthogonal (GOE) or Symplectic Ensemble (GSE), respectively, while the Gaussian Unitary Ensemble (GUE) describes the fluctuation properties of a system under broken time reversal invariance. These ensembles provide a generic description of chaotic spectral fluctuations. However, regular fluctuation properties are not generic in this sense, they rather differ from system to system. One example is the harmonic oscillator whose energy levels are equidistant. Our main interest in this work will be the opposite situation, i.e. a regular system that does not show any spectral correlations. Such a system is then modeled by the Poisson Ensemble.

Statistical models of this type can also be employed to understand similar transitions in other systems. Classical and quantum billiards [4] are very useful systems for chaos studies because of their relatively small number of degrees of freedom. Starting from a regularly shaped billiard, slight changes of the geometry often make the dynamics gradually or even abruptly chaotic. The Hénon-Heiles system [4] can be viewed as the truncation of a regular and integrable system such that the transition from regular to chaotic behavior occurs as energy increases. Similar observations have been made in molecular spectra. Hence, the decomposition into a regular and a chaotic part is not always dictated by externally tunable parameters like magnetic field or deformations of the geometry. Sometimes, a very complicated system becomes mathematically feasible if written as a sum of a somehow smooth, dominating term and a correction. The nuclear shell model [6] might be among the most successful examples for the utility of this idea. The potential is a sum of an usually integrable mean field part and an often chaos producing residual interaction. Another example from nuclear physics is provided by heavy ion reactions where compound nuclei are produced with very high angular momentum. Their rotational motion along the yrast line is collective and integrable. However, often there is an inner temperature due to single particle excitations creating chaos which becomes visible in a spreading of the electrical quadrupole transition strength [7].

Recently, it has been shown that the statistical random matrix model involving a sum of a regular and a chaotic part is even useful to analyze spectra in classical wave mechanics. The spectral fluctuations of elastomechanical eigenmodes in Aluminum blocks manufactured in the shape of three dimensional Sinai-billiards show a transition which closely parallels the transition from Poisson regularity to chaos in quantum mechanics [8].

Obviously, the detailed study of the transition from regularity, in particular Poisson regularity, to chaos in such a random matrix model is a worthwhile task. Of the numerous studies addressing this problem, we mention particularly the work of French, Kota, Pandey and Tomsovic [9] and the work of Leyvraz and Seligman [10]. In both studies, the two level correlation function for small values of the transition parameter is derived. Detailed numerical simulations can be found in Ref. [7]. However, despite several attempts, a full fledged analytical discussion of this transitions is still lacking. The calculation that comes closest to a solution of the mathematical problems is most likely the one presented by Lenz [11]. Surprisingly, other but related transitions like the breaking of time reversal invariance could be calculated analytically. In a statistical model, the Hamiltonian of a system with gradually broken time reversal invariance is expressible as a sum of two contributions, one drawn from the GOE, the other from the GUE. The spectral fluctuations of this transition were worked out analytically by Mehta and Pandey [12]. Using the level number variance as a measure, upper bounds for the time reversal invariance breaking in nuclear physics were given [13].

In this work, we study transitions from arbitrary to chaotic fluctuation properties. Our main interest is the transition from Poisson regularity to chaos, but many of our results are very general and include this transition and the time reversal invariance breaking as special cases. The only general restriction we make is that the Hamiltonian can be written as the sum of two parts such that one of them can be modeled by a Gaussian Ensemble. Although most of the physically interesting transitions fall into this class, there are exceptions. The breaking of a quantum number, for example, has of course a completely different Hamiltonian structure. The influence of isospin breaking on the fluctuation properties has been studied experimentally [14] and theoretically [15] in great detail. Those systems are not the subject of this work. Moreover, for technical reasons, most of our results involving integral representations are derived for transitions toward the GUE.

For the mathematical discussion of the transition we rely on supersymmetry. In Refs. [16, 17] supersymmetric nonlinear $\sigma$-models were derived for a wide class of statistical quantum systems with chaotic dynamics. In Refs. [15, 18] a particularly suitable technique was developed in order to treat these models in the pure spectral case.

In Section 2, we formulate the random matrix model and introduce our notation. Supersymmetric methods are applied in Section 3. In Section 4, we derive diffusion equations for the transition. The transition from Poisson regularity to chaos is worked out in Section 5. We discuss our results in Section 6. Certain calculations are performed in the Appendix.

## 2 Transitions toward the GUE

In Section 2.1, we summarize the main results for pure ensembles and introduce our notation. In Section 2.2 the model for the transition is introduced. For simplicity, all formulae are derived for the GUE. With the aid of Ref. [3], however, everything can easily be mapped onto the GOE or the GSE.

### 2.1 Pure Ensembles

We consider two ensembles of Hermitean $N \times N$ matrices $H^{(0)}$ and $H^{(1)}$, respectively, with volume elements

$$d[H^{(j)}] = \prod_{n=1}^{N} dH_{nn}^{(j)} \prod_{n>m} d\mathrm{Re}H_{nm}^{(j)} d\mathrm{Im}H_{nm}^{(j)} \ , \quad j = 0, 1 \tag{2.1}$$

and normalized probability density functions $P_N^{(j)}(H^{(j)})$, $j = 0, 1$. Some well known formulae which are true for arbitrary ensembles are given in the following. The $k$-level correlation functions

$$R_k^{(j)}(x_1, \ldots, x_k) = \frac{1}{\pi^k} \int P_N^{(j)}(H^{(j)}) \prod_{p=1}^{k} \mathrm{Im\,tr} \frac{1}{x_p^- - H^{(j)}} d[H^{(j)}] \tag{2.2}$$

measure the probability of finding a level in each of the unit intervals around the energies $x_1, \ldots, x_k$. The energies carry a small imaginary increment, $x_p^- = x_p - i\varepsilon$, the limit $\varepsilon \to 0$ has to be taken after the ensemble average. Notice that this definition differs slightly from the Dyson-Mehta definition [3] since it includes terms proportional to $\delta(x_p - x_q)$. We are not interested in those contributions and will therefore proceed as in Refs. [15, 18] by always assuming $x_p \neq x_q$ for all $p \neq q$. In order to study the fluctuations, the correlation functions have to be unfolded by measuring the energies on the scale of the mean level spacing $D^{(j)} = 1/R_1^{(j)}(0)$, $j = 0, 1$ for infinitely many levels. We define the dimensionless energies $\xi_p = x_p/D^{(j)}$, $p = 1, \ldots, k$ which have to be held fixed in the limit $N \to \infty$. The unfolded $k$-level correlation functions

$$X_k^{(j)}(\xi_1, \ldots, \xi_k) = \lim_{N \to \infty} (D^{(j)})^k R_k^{(j)}(x_1, \ldots, x_k) \qquad (2.3)$$

measure the correlations on the energy scale with mean level spacing unity everywhere. All physically important ensembles allow such a complete removal of the level densities from the higher correlations. The functions $X_k^{(j)}(\xi, \ldots, \xi_k)$ depend then only on the differences $\xi_p - \xi_q$. Those fluctuation properties are called translation invariant or generic.

By including the real part of the Green function, we introduce the correlation functions

$$\widehat{R}_k^{(j)}(x_1, \ldots, x_k) = \frac{1}{\pi^k} \int P_N^{(j)}(H^{(j)}) \prod_{p=1}^k \operatorname{tr} \frac{1}{x_p^\pm - H^{(j)}} d[H^{(j)}] \qquad (2.4)$$

where the imaginary increments can now lie on both sides of the real axis, we write $x_p^\pm = x_p \pm i\varepsilon$. Due to the identity

$$\operatorname{Im} \operatorname{tr} \frac{1}{x_p^- - H^{(j)}} = \frac{1}{i2} \left( \operatorname{tr} \frac{1}{x_p^- - H^{(j)}} - \operatorname{tr} \frac{1}{x_p^+ - H^{(j)}} \right) , \qquad (2.5)$$

the physically relevant correlations (2.2) can always be constructed as a linear combination of the functions (2.4). As in Ref. [15], we use the operator symbol $\Im$ in order to indicate this procedure. The functions (2.4) have the advantage of being expressible as the derivatives

$$\widehat{R}_k^{(j)}(x_1, \ldots, x_k) = \frac{1}{(2\pi)^k} \frac{\partial^k}{\prod_{p=1}^k \partial J_p} Z_k^{(j)}(x + J) \bigg|_{J=0} \qquad (2.6)$$

of the generating function

$$Z_k^{(j)}(x + J) = \int P_N^{(j)}(H^{(j)}) \prod_{p=1}^k \frac{\det\left(x_p^\pm + J_p - H^{(j)}\right)}{\det\left(x_p^\pm - J_p - H^{(j)}\right)} d[H^{(j)}] \qquad (2.7)$$

with source variables $J_p$, $p = 1, \ldots, k$. For later purposes we write the argument as $x + J$ with the $2k \times 2k$ matrices

$$\begin{aligned} x &= \operatorname{diag}(x_1, x_1, \ldots, x_k, x_k) \\ J &= \operatorname{diag}(-J_1, +J_1, \ldots, -J_k, +J_k) . \end{aligned} \qquad (2.8)$$

The generating function is normalized, i.e. at $J = 0$ we have $Z_k^{(j)}(x) = 1$. The function $\Im Z_k^{(j)}(x + J)$ generates the correlations (2.2). We emphasize again that all these formulae presented so far are true for arbitrary probability densities.

By choosing a Gaussian distribution for the matrices $H^{(1)}$,

$$P_N^{(1)}(H^{(1)}) = \frac{2^{N(N-1)/2}}{(2\pi v^2)^{N^2/2}} \exp\left(-\frac{1}{2v^2}\text{tr}(H^{(1)})^2\right) \quad (2.9)$$

with variance $v^2$, we make one of the ensembles a Gaussian Unitary Ensemble (GUE) with chaotic fluctuation properties. One usually fixes the energy scale by setting $\sqrt{2}v = 1$. We have the well known result [3]

$$\begin{aligned} R_k^{(1)}(x_1,\ldots,x_k) &= \det[K_N(x_p,x_q)]_{p,q=1,\ldots,k} \\ K_N(x_p,x_q) &= \sum_{n=0}^{N-1} \varphi_n(x_p)\varphi_n(x_q) \end{aligned} \quad (2.10)$$

where $\varphi_n(x_p)$ is the $n$-th oscillator wave function. The lower order correlations are removed in the $k$-level cluster functions

$$T_k^{(1)}(x_1,\ldots,x_k) = \sum_G (-1)^{m-k}(m-1)! \prod_{m'=1}^{m} R_{G_{m'}}^{(1)}, \quad (2.11)$$

where $G$ denotes a partition of the indices $(1,\ldots,k)$ into subgroups $(G_1,\ldots,G_m)$. The inverse reads

$$R_k^{(1)}(x_1,\ldots,x_k) = \sum_G (-1)^{k-m} \prod_{m'=1}^{m} T_{G_{m'}}^{(1)}. \quad (2.12)$$

Besides the unfolded $k$-level correlation functions (2.3) we also define the unfolded $k$-level cluster functions

$$Y_k^{(1)}(\xi_1,\ldots,\xi_k) = \lim_{N\to\infty} (D^{(1)})^k T_k^{(1)}(x_1,\ldots,x_k). \quad (2.13)$$

The other probability density function $P_N^{(0)}(H^{(0)})$ we do not specify and thereby allow for arbitrary fluctuation properties.

## 2.2 Transition Ensemble

We want to study the fluctuations of the ensemble of matrices

$$H(\alpha) = H^{(0)} + \alpha H^{(1)} \quad (2.14)$$

as a function of the transition parameter $\alpha$. In this formulation, the arbitrary ensemble is the initial condition, and as we increase $\alpha$, the GUE fluctuation properties will become stronger and eventually dominate. It is advantageous to absorb the transition parameter into the GUE by defining $\alpha H^{(1)}$ as new matrices. This can formally be achieved by replacing $\sqrt{2}v$ in equation (2.9) with $\alpha$ such that $\alpha^2/2$ is now the variance of the Gaussian distribution. We write $P_N^{(1)}(H^{(1)}, \alpha)$ to indicate the dependence on the transition parameter. At the initial condition, the Gaussian distribution becomes a $\delta$-function,

$$\lim_{\alpha \to 0} P_N^{(1)}(H^{(1)}, \alpha) = \delta(H^{(1)}) . \qquad (2.15)$$

The $k$-level correlation functions of this transition ensemble are given by

$$R_k(x_1, \ldots, x_k, \alpha) = \frac{1}{\pi^k} \int d[H^{(0)}] P_N^{(0)}(H^{(0)}) \int d[H^{(1)}] P_N^{(1)}(H^{(1)}, \alpha)$$
$$\prod_{p=1}^{k} \operatorname{Im} \operatorname{tr} \frac{1}{x_p^- - H^{(0)} - H^{(1)}} \qquad (2.16)$$

and the $k$-level cluster functions $T_k(x_1, \ldots, x_k, \alpha)$ are defined as in Eq. (2.11). Again, easier to evaluate are the functions

$$\widehat{R}_k(x_1, \ldots, x_k, \alpha) = \frac{1}{\pi^k} \int d[H^{(0)}] P_N^{(0)}(H^{(0)}) \int d[H^{(1)}] P_N^{(1)}(H^{(1)}, \alpha)$$
$$\prod_{p=1}^{k} \operatorname{tr} \frac{1}{x_p^- - H^{(0)} - H^{(1)}} \qquad (2.17)$$

since they can be written as derivatives

$$\widehat{R}_k(x_1, \ldots, x_k, \alpha) = \frac{1}{(2\pi)^k} \frac{\partial^k}{\prod_{p=1}^{k} \partial J_p} Z_k(x + J, \alpha) \bigg|_{J=0} \qquad (2.18)$$

of the normalized generating function

$$Z_k(x + J, \alpha) = \int d[H^{(0)}] P_N^{(0)}(H^{(0)}) \int d[H^{(1)}] P_N^{(1)}(H^{(1)}, \alpha)$$
$$\prod_{p=1}^{k} \frac{\det\left(x_p^\pm + J_p - H^{(0)} - H^{(1)}\right)}{\det\left(x_p^\pm - J_p - H^{(0)} - H^{(1)}\right)} . \qquad (2.19)$$

Again, the physically relevant correlations (2.16) are generated by the function $\Im Z_k(x + J, \alpha)$. Because of Eq. (2.15) we see immediately that the correlations of the arbitrary ensemble emerge for vanishing transition parameter,

$$\lim_{\alpha \to 0} R_k(x_1, \ldots, x_k, \alpha) = R_k^{(0)}(x_1, \ldots, x_k)$$
$$\lim_{\alpha \to 0} Z_k(x + J, \alpha) = Z_k^{(0)}(x + J) . \qquad (2.20)$$

The limit $\alpha \to \infty$ has to yield the GUE result. In order to check that, we do not absorb $\alpha$ into $H^{(1)}$ but use the Hamiltonian in the form (2.14), we find

$$R_k(x_1,\ldots,x_k,\alpha) = \frac{1}{(\alpha\pi)^k} \int d[H^{(0)}]\, P_N^{(0)}(H^{(0)}) \int d[H^{(1)}]\, P_N^{(1)}(H^{(1)},1)$$

$$\prod_{p=1}^{k} \operatorname{Im} \operatorname{tr} \frac{1}{x_p^-/\alpha - H^{(0)}/\alpha - H^{(1)}} \ . \qquad (2.21)$$

Obviously, the energies have to be rescaled when we compute the limit $\alpha \to \infty$ by holding fixed $\widetilde{x}_p = x_p/\alpha$. The factor $1/\alpha^k$ in front of the integral takes care of the proper rescaling of the differentials,

$$\lim_{\alpha \to \infty} R_k(x_1,\ldots,x_k,\alpha)\, dx_1 \cdots dx_k = R_k^{(1)}(\widetilde{x}_1,\ldots,\widetilde{x}_k)\, d\widetilde{x}_1 \cdots d\widetilde{x}_k \qquad (2.22)$$

as required. Equivalently we can write

$$\lim_{\alpha \to \infty} \alpha^k\, R_k(x_1,\ldots,x_k,\alpha) = R_k^{(1)}(\widetilde{x}_1,\ldots,\widetilde{x}_k) \qquad (2.23)$$

and similarly for the generating function.

In the definition (2.14) of the transition ensemble, the parameter $\alpha$ can take values between zero and infinity. Since the ensemble is obviously symmetric in the transition parameter, we consider only non-negative values of $\alpha$. In certain situations, however, it is more convenient to choose alternative definitions. In general, we can write

$$H(\alpha) = \gamma^{(0)}(\alpha)H^{(0)} + \gamma^{(1)}(\alpha)H^{(1)} \qquad (2.24)$$

where $\gamma^{(0)}(\alpha)$ should be symmetric and $\gamma^{(1)}(\alpha)$ antisymmetric in $\alpha$. A good example is $\gamma^{(0)}(\alpha) = \sqrt{1-\alpha^2}$ and $\gamma^{(1)}(\alpha) = \alpha$. Here, $\alpha$ varies only between zero and unity. One verifies easily that the limits $\alpha \to 0$ and $\alpha \to 1$ give the correct results for the arbitrary ensemble and for the GUE, respectively. Another choice is $\gamma^{(0)}(\alpha) = \cos\alpha$ and $\gamma^{(1)}(\alpha) = \sin\alpha$, where $\alpha$ varies between zero and $\pi/2$. We will mainly work with the definition (2.14).

As in the case of the pure ensembles, we have to unfold the correlation functions for large level number $N$ by removing the dependence on the level density. We measure the fluctuations on the scale of the mean level spacing $D = 1/R_1(0,\alpha)$, i.e. we map the spectrum onto an equivalent one with mean level spacing unity everywhere. Obviously, this can make sense only if the level densities of the pure ensembles involved cover the same section of the energy scale $x$. In other words, $R_1^{(0)}(x)$ and $R_1^{(1)}(x)$ should have the same asymptotic dependence on the level number. The GUE level density $R_1^{(1)}(x)$ becomes asymptotically the famous Wigner semicircle [3] which behaves as $\sqrt{N}$. Thus, the mean level spacing is proportional to $1/\sqrt{N}$. We define new energies $\xi_p = x_p/D$, $p = 1, \ldots, k$. The transition parameter $\alpha$ is defined on the original scale and has therefore also to be unfolded. The new transition parameter

$$\lambda = \alpha/D \qquad (2.25)$$

was first introduced by Pandey [19], its universality was shown in many applications [7, 9, 12, 15]. Since $\lambda$ has to be held fixed in the limit $N \to \infty$, the original transition parameter $\alpha$ behaves formally as $1/\sqrt{N}$, just like the mean level spacing. The $k$-level correlation functions on the unfolded scale

$$X_k(\xi_1, \ldots, \xi_k, \lambda) = \lim_{N \to \infty} D^k R_k(x_1, \ldots, x_k, \alpha) \qquad (2.26)$$

are translation invariant. By including the real parts of the Green function, we also define the functions $\widehat{X}_k(\xi, \ldots, \xi_k, \lambda)$ for later purposes.

## 3 Application of Supersymmetry

In Section 3.1 the supersymmetric representation is introduced. The graded eigenvalue method is applied in Section 3.2. In Section 3.3 a Pastur equation for the level density is derived. The unfolding procedure is performed in Section 3.4. In Section 3.5, the correlation functions on the unfolded energy scale are expanded in the transition parameter. Translation invariance and its consequences are discussed in Section 3.6. In Section 3.7, the level number variance is worked out.

### 3.1 Supersymmetric Representation

The average over the GUE in the generating function (2.19) can be evaluated by using the supersymmetric technique developed in Refs. [16, 17]. The advantage of this technique is a dramatic reduction of the degrees of freedom which becomes obvious in the decoupling of the number of integrations to be performed from the level number $N$. The procedure can straightforwardly be applied to our case and will therefore not be discussed here. Since we will later not employ the coset method of Refs. [16, 17], the reader who is not familiar with supersymmetry in this context is advised to study Ref. [18] which also defines the notation for the remainder.

The GUE is mapped onto an ensemble of $2k \times 2k$ Hermitean graded or supermatrices $\sigma$. In $pq$-block notation [17, 18], these matrices are defined as $k \times k$ matrices whose entries $\sigma_{pq}$ are $2 \times 2$ supermatrices of the form

$$\sigma_{pq} = \begin{bmatrix} \sigma_{pq}^{C1} & \sigma_{pq}^{A*} \\ \sigma_{pq}^{A} & i\sigma_{pq}^{C2} \end{bmatrix} \qquad (3.1)$$

where $\sigma_{pq}^{Cc}, c = 1, 2$ and $\sigma_{pq}^{A}$ are commuting and anticommuting variables, respectively. The Cartesean volume element reads

$$d[\sigma] = \prod_{c=1}^{2} \prod_{p=1}^{k} d\sigma_{pp}^{Cc} \prod_{p>q} d\mathrm{Re}\sigma_{pq}^{Cc} d\mathrm{Im}\sigma_{pq}^{Cc} \prod_{p,q} d\sigma_{pq}^{A*}\sigma_{pq}^{A} . \tag{3.2}$$

There is a normalized graded probability density function

$$Q_k(\sigma, \alpha) = 2^{k(k-1)} \exp\left(-\frac{1}{\alpha^2}\mathrm{trg}\sigma^2\right) \tag{3.3}$$

having the well defined limit

$$\lim_{\alpha \to 0} Q_k(\sigma, \alpha) = \delta(\sigma) \tag{3.4}$$

where the right hand side is a properly defined $\delta$-function in superspace [16, 18]. The generating function (2.19) can then be expressed in the form

$$Z_k(x + J, \alpha) = \int d[H^{(0)}] P_N^{(0)}(H^{(0)}) \int d[\sigma] Q_k(\sigma, \alpha)$$

$$\mathrm{detg}^{-1}\left((x^{\pm} + J - \sigma) \otimes 1_N - 1_{2k} \otimes H^{(0)}\right) \tag{3.5}$$

in which we introduced the $N \times N$ and the $2k \times 2k$ unit matrices $1_N$ and $1_{2k}$. Notice that the order $k$ of the correlation is now directly reflected in the dimension $2k$ of the supermatrix $\sigma$, this causes the enormous simplification. For the two-level correlation function of the Gaussian Orthogonal Ensemble (GOE), a similar integral representation of the generating function was obtained in Ref. [7].

## 3.2 Graded Eigenvalue Method

In order to further evaluate the generating function we use the graded eigenvalue method which was developed in Ref. [18]. After a simple shift of variables and a interchange of the order of integration we find

$$Z_k(x + J, \alpha) = \int d[\sigma] Q_k(\sigma^{\mp} - x - J, \alpha) \int d[H^{(0)}] P_N^{(0)}(H^{(0)})$$

$$\mathrm{detg}^{-1}\left(\sigma \otimes 1_N - 1_{2k} \otimes H^{(0)}\right) \tag{3.6}$$

such that due to the direct product structure of the argument the superdeterminant depends only on the eigenvalues of $\sigma$ and $H^{(0)}$. All angular degrees of freedom are now in the graded probability density function. We therefore go over to eigenvalue and angle coordinates

$$\sigma = u^{-1}su \tag{3.7}$$

where $u$ is an element of the unitary supergroup $U(k/k)$. The diagonal matrix of the eigenvalues
$$s = \mathrm{diag}(s_{11}, is_{12}, \ldots, s_{k1}, is_{k2}) \tag{3.8}$$
contains $k$ eigenvalues $s_{p1}$ of boson boson and as many eigenvalues $is_{p2}$ of fermion fermion type [17, 18].

For the integration over the supermatrix, we need the transformation of the Cartesean volume element (3.2) to eigenvalue-angle coordinates [18],
$$d[\sigma] = B_k^2(s) \, d[s] \, d\mu(u) \,, \tag{3.9}$$
where $d[s]$ is the product of the differentials of the eigenvalues and $d\mu(u)$ is the invariant Haar measure of the unitary supergroup. The square root of the Jacobian, here referred to as Berezinian, is given by
$$B_k(s) = \frac{\prod_{p>q}^k (s_{p1} - s_{q1}) \prod_{p>q}^k (is_{p2} - is_{q2})}{\prod_{p,q} (s_{p1} - is_{q2})} = \det \left[ \frac{1}{s_{p1} - is_{q2}} \right]_{p,q=1,\ldots,k} \tag{3.10}$$
which reflects [18] the determinant structure (2.10) of the GUE correlation functions.

The angular average over the unitary supergroup can be performed by using the supersymmetric generalization [18] of the Harish-Chandra-Itzykson-Zuber integral [20, 21]. We define a second Hermitean supermatrix $v^{-1}rv$ of the type (3.7) where $v$ is superunitary and $r$ diagonal. The angular average over the shifted Gaussian probability density function yields
$$\int Q_k(u^{-1}su - v^{-1}rv, \alpha) \, d\mu(u) = \frac{G_k(s, r, \alpha)}{B_k(s) B_k(r)} \tag{3.11}$$
where the Gaussian kernel on the right hand side is given by
$$G_k(s, r, \alpha) = \det [g(s_{p1} - r_{q1}, \alpha)]_{p,q=1,\ldots,k} \det [g(s_{p2} - r_{q2}, \alpha)]_{p,q=1,\ldots,k}$$
$$g(z, \alpha) = \frac{1}{\sqrt{\pi \alpha^2}} \exp\left(-\frac{z^2}{\alpha^2}\right) \,. \tag{3.12}$$

This result does not depend on the superunitary matrix $v$ because of the invariance of the Haar measure $d\mu(u)$. Since the determinants and the Berezinian (3.10) are antisymmetric, simple interchanges of variables show that all terms in the Laplace expansion of $G_k(s, r, \alpha)$ contribute the same if further integration over the eigenvalues $s$ or $r$ is required and the rest of the integrand has the same symmetries. Hence, in our case, it is sufficient to write
$$G_k(s, r, \alpha) = G_k(s - r, \alpha) = \frac{1}{\sqrt{\pi \alpha^2}^{2k}} \exp\left(-\frac{1}{\alpha^2} \mathrm{trg}(s - r)^2\right) \tag{3.13}$$

under the integral. Furthermore, the integration over $s$ requires to take a new type of boundary contributions [16, 17, 22, 23] into account which do not occur in ordinary analysis. Following Ref. [18] we have to add the term $(1 - \eta(r))\delta(s)/B_k^2(s)$ to the right hand side of the formula (3.11) which contains the distribution

$$\eta(r) = \begin{cases} 0 & \text{if } r_{p1} = ir_{q2} \text{ for any } p, q \\ 1 & \text{otherwise} \end{cases} . \tag{3.14}$$

In our particular case, we want to apply formula (3.11) to the integral (3.6). Thus, we have a diagonal shift matrix, i.e. $v = 0$, with eigenvalues given by $r_{p1} = x_p \pm i\varepsilon - J_p$ and $ir_{p2} = x_p \pm i\varepsilon + J_p$. It is then obvious from Refs. [18, 24] that formula (3.11) is valid for all choices of the signs of the imaginary increments. Consequently, the procedure to go over to the correlations of the imaginary parts which was designated by the operator symbol $\Im$ does still work.

After performing the angular integration we shift the imaginary increments back into the superdeterminant. A comparison with the definition (2.7) shows that the use of the eigenvalue angle coordinates allows to express the average over the arbitrary ensemble as nothing else but the generating function of the arbitrary correlations,

$$Z_k^{(0)}(s) = \int P_N^{(0)}(H^{(0)}) \det g^{-1}\left(s^\pm \otimes 1_N - 1_{2k} \otimes H^{(0)}\right) d[H^{(0)}] \tag{3.15}$$

where $s$ takes the place of $x + J$. Collecting everything we arrive at

$$Z_k(x+J, \alpha) = 1 - \eta(x+J) + \frac{1}{B_k(x+J)} \int G_k(s - x - J, \alpha) Z_k^{(0)}(s) B_k(s) d[s] \tag{3.16}$$

for the generating function of the transition ensemble. The boundary contribution ensures the normalization at $J = 0$ such that $Z_k(x, \alpha) = 1$. However, in the limit of vanishing transition parameter $\alpha$, the Gaussian kernel (3.12) becomes a $\delta$-function, in particular we find the diagonal term $\delta(s - x - J)$ for the expression (3.13). The integral yields then the initial condition $Z_k^{(0)}(x+J)$ for all values of the source variables, including $J = 0$ in perfect agreement with Eq. (2.20). Hence, in this special case, the boundary contribution $1 - \eta(x+J)$ is obsolete. We keep that in mind, but we do not indicate that through a special notation. There is still another caveat in order. It was shown in Ref. [25] that the treatment of the boundary terms which is used here can miss some contributions to the real parts of the correlation functions. Nevertheless, since everything becomes correct in the physically relevant limit of infinitely many levels and since we are only interested in the imaginary parts anyway, we can proceed with formula (3.16).

It is now very easy to go over from the generating function (3.16) to the $k$-level correlation functions. For $\alpha = 0$ we find by construction

$$\widehat{R}_k(x_1, \ldots, x_k, 0) = \widehat{R}_k^{(0)}(x_1, \ldots, x_k) \tag{3.17}$$

which is consistent with Eq. (2.20). For non-zero $\alpha$, we arrive at

$$\widehat{R}_k(x_1, \ldots, x_k, \alpha) = \frac{(-1)^k}{\pi^k} \int G_k(s - x, \alpha) Z_k^{(0)}(s) B_k(s) d[s] \qquad (3.18)$$

by using the methods of Ref. [18]. In order to find the correlations of the imaginary parts we simply have to replace $Z_k^{(0)}(s)$ in the integrand of Eqs. (3.16) and (3.18) with $\Im Z_k^{(0)}(s)$. The $k$-level correlation function for the transition from arbitrary initial conditions toward the GUE is represented as a $2k$-fold integral. Formula (3.11) is crucial for the graded eigenvalue method. In a single step, all angular degrees of freedom including all Grassmann variables are integrated out. This is true for arbitrary index $k$ of the correlations. In the coset method used in Refs. [16, 17], the evaluation of correlations with indices $k > 2$ becomes increasingly complicated.

## 3.3 Level Density and Pastur Equation

The graded eigenvalue method is especially suited for a discussion of the correlations in the case of transitions [15]. However, the asymptotic behavior of the level density for large level numbers can easier be found by using the saddlepoint method of Refs. [16, 17]. Starting from Eq. (3.6) and using Eq. (3.15) we can write

$$Z_1(x + J, \alpha) = \int \exp L(\sigma, x + J, \alpha) d[\sigma] \qquad (3.19)$$

with the Lagrange function

$$L(\sigma, x + J, \alpha) = -\frac{1}{\alpha^2} \text{trg}(\sigma - x - J)^2 + \ln Z_1^{(0)}(s) . \qquad (3.20)$$

Since at the saddlepoint we can set $J = 0$, the saddlepoint value $\sigma = \sigma_0$ is determined through the equation

$$0 = dL(\sigma, x, \alpha)\bigg|_{\sigma=\sigma_0} = \text{trg}\left[-\frac{2}{\alpha^2}(\sigma - x) + \frac{\partial}{\partial \sigma} \ln Z_1^{(0)}(s)\right]_{\sigma=\sigma_0} d\sigma \qquad (3.21)$$

which is solved in Appendix A. The solution is diagonal, both eigenvalues are equal, $s_{10} = is_{20}$, and satisfy the implicit equation

$$\frac{2}{\alpha^2}(s_{10} - x) + \pi \widehat{R}_1^{(0)}(s_{10}) = 0 \qquad (N \to \infty) \qquad (3.22)$$

where we omit the index $k = 1$ in the eigenvalues and the energy. Here, of course, $\widehat{R}_1^{(0)}(s_{10})$ stands for the leading term in an asymptotic expansion for large level number $N$.

The asymptotic form of the one point function is then found to be

$$\widehat{R}_1(x, \alpha) = \frac{2}{\pi \alpha^2} (x - s_{10}(x, \alpha)) = \widehat{R}_1^{(0)}(s_{10}(x, \alpha)) \qquad (3.23)$$

where $s_{10} = s_{10}(x,\alpha)$ is the solution of the implicit equation (3.22), details are sketched in Appendix A. The level density

$$R_1(x,\alpha) = -\frac{2}{\pi\alpha^2}\mathrm{Im}\, s_{10}(x,\alpha) = \mathrm{Im}\,\widehat{R}_1^{(0)}(s_{10}(x,\alpha)) \qquad (3.24)$$

is the imaginary part of the one point function. All these results have to be understood asymptotically, i.e. for large level number $N$. The level density $R_1(x,\alpha)$ is always normalized if the arbitrary initial condition is normalized to the total level number,

$$\begin{aligned}
\int_{-\infty}^{+\infty} R_1(x,\alpha)\, dx &= \int_{-\infty}^{+\infty} R_1^{(0)}(s_{10}(x,\alpha))\, dx \\
&= \int_{-\infty}^{+\infty} R_1^{(0)}(s_{10})\, ds_{10} + \frac{\pi\alpha^2}{2}\mathrm{Im}\int_{-\infty}^{+\infty} \widehat{R}_1^{(0)}(s_{10})\frac{d\widehat{R}_1^{(0)}}{ds_{10}}(s_{10})\, ds_{10} \\
&= N\,. \qquad (3.25)
\end{aligned}$$

We changed the variable of integration from $x$ to $s_{10}$ with the help of the Eqs. (3.22) and (3.23). The second integral in the second line is zero since for consistency reasons the level density $R_1^{(0)}(x)$ has to vanish for infinitely large energy argument.

Pastur [26] derived the implicit equation (3.22) without using supersymmetry. It appears here through a simple saddlepoint approximation because supersymmetry decouples the level number $N$ from the integration. The implicit equation for the pure Gaussian ensembles can easily be obtained by choosing the unaveraged Green function

$$\widehat{R}_1^{(0)}(s_{10}) = \frac{N}{\pi}\frac{1}{s_{10}} \qquad (3.26)$$

as initial condition and by setting $\alpha = 1$, we find from Eq. (3.22)

$$2(s_{10} - x) + \frac{N}{s_{10}} = 0 \qquad (3.27)$$

which yields the well-known semicircle law. More interesting, however, is the case of having a semicircle as initial condition,

$$\widehat{R}_1^{(0)}(s_{10}) = \frac{1}{\pi}\left(s_{10} + i\sqrt{2N - s_{10}^2}\right)\,. \qquad (3.28)$$

We solve equation (3.22) for $s_{10}$ and find from formula (3.23) another semicircle law

$$\widehat{R}_1(x,\alpha) = \frac{1}{\pi\sqrt{1+\alpha^2}}\left(\frac{x}{\sqrt{1+\alpha^2}} + i\sqrt{2N - \left(\frac{x}{\sqrt{1+\alpha^2}}\right)^2}\right) \qquad (3.29)$$

which involves the rescaled energy $x/\sqrt{1+\alpha^2}$. Note that this result is correct for all values of $\alpha$.

## 3.4 Unfolding Procedure

The main advantage due to the application of supersymmetry and the graded eigenvalue method is the decoupling of the degrees of freedom of the supermatrices from the level number which appears only in the initial condition $Z_k^{(0)}(s)$. This feature makes it possible in many situations to perform the unfolding procedure in the generating functions. Besides the new energies $\xi_p = x_p/D$ we also define new source variables $j_p = J_p/D$, $p = 1, \ldots, k$ and the $2k \times 2k$ matrices $\xi$ and $j$ analogously to Eq. (2.8). The $k$-level correlation functions on the unfolded energy scale depending on the transition parameter $\lambda$ can then be written as the $k$-fold derivative

$$\widehat{X}_k(\xi_1, \ldots, \xi_k, \lambda) = \frac{1}{(2\pi)^k} \frac{\partial^k}{\prod_{p=1}^k \partial j_p} z_k(\xi + j, \lambda) \bigg|_{j=0} \quad (3.30)$$

of the unfolded generating function

$$z_k(\xi + j, \lambda) = \lim_{N \to \infty} Z_k(x + J, \alpha) \ . \quad (3.31)$$

It is useful to unfold the integration variables $s$ in Eq. (3.16) by making the change of variables $s \to s/D$. We find

$$z_k(\xi + j, \lambda) = 1 - \eta(\xi + j) + \frac{1}{B_k(\xi + j)} \int G_k(s - \xi - j, \lambda) z_k^{(0)}(s) B_k(s) d[s] \quad (3.32)$$

where the unfolded initial condition is given by

$$z_k^{(0)}(s) = \lim_{N \to \infty} Z_k^{(0)}(Ds) \ . \quad (3.33)$$

Consequently, the whole unfolding procedure is naturally shifted into the initial condition. As we will see in the following two sections, this is very useful from a conceptual viewpoint and also for applications. Analogous to Eqs. (3.17) and (3.18), we find for the $k$-level correlation functions on the unfolded scale

$$\widehat{X}_k(\xi_1, \ldots, \xi_k, 0) = \widehat{X}_k^{(0)}(\xi_1, \ldots, \xi_k) \quad (3.34)$$

for $\lambda = 0$ and by using Eq. (3.30) we have

$$\widehat{X}_k(\xi_1, \ldots, \xi_k, \lambda) = \frac{(-1)^k}{\pi^k} \int G_k(s - \xi, \lambda) z_k^{(0)}(s) B_k(s) d[s] \quad (3.35)$$

for non-zero $\lambda$. As on the original scale, we simply have to replace $z_k^{(0)}(s)$ by $\Im z_k^{(0)}(s)$ in order to go over to the physically interesting contributions from the imaginary parts of the Green functions.

A comment on the alternative definitions (2.24) of the transition parameter $\alpha$ is in order. The parameterization $\gamma^{(0)}(\alpha) = \sqrt{1-\alpha^2}$ and $\gamma^{(1)}(\alpha) = \alpha$ gives on the unfolded scale the same result (3.32) since $\alpha$ behaves formally as $1/\sqrt{N}$ and thus $\lim_{N\to\infty} \gamma^{(0)}(\alpha) = 1$. This is also true for the choice $\gamma^{(0)}(\alpha) = \cos\alpha$ and $\gamma^{(1)}(\alpha) = \sin\alpha$. We conclude that a wide class of parameterizations yields the same result (3.32) on the unfolded scale. The limit $\lambda \to 0$ gives the fluctuations of the arbitrary initial condition. The GUE results have to show up for $\lambda \to \infty$. However, it is known from several applications [7, 9, 12, 15] that the chaotic features start dominating the fluctuations for much smaller values of $\lambda$. In most cases, a value of $\lambda$ around unity will be enough to spread the GUE correlations over the spectrum.

## 3.5 Expansion in the Transition Parameter

The $k$-level correlation functions $X_k(\xi_1,\ldots,\xi_k,\lambda)$ following from Eq. (3.35) have a remarkable scaling property which shows up when the variables $s/\lambda$ are introduced for the integration,

$$X_k(\xi_1,\ldots,\xi_k,\lambda) = \frac{(-1)^k}{\pi^k} \int G_k(s-\xi/\lambda, 1) \frac{\Im z_k^{(0)}(\lambda s)}{\lambda^k} B_k(s) d[s] . \qquad (3.36)$$

The transition parameter $\lambda$ has disappeared from the variance of the Gaussian function and occurs only in the initial condition and in the combination $\xi/\lambda$. Since this rescaling of the energies is uncritical for all parameter values, we may hold the ratios $\xi_p/\lambda, p = 1,\ldots,k$ fixed and expand the initial condition in the transition parameter. Two boundary conditions are present: firstly, the correlations must obviously be an even function in $\lambda$ and secondly, there should not be a singularity for vanishing $\lambda$. Hence, it is reasonable to assume an expansion of the form

$$\frac{\Im z_k^{(0)}(\lambda s)}{\lambda^k} = \sum_{n=0}^{\infty} \frac{\lambda^{2n}}{n!} w_{k,n}^{(0)}(s) \qquad (3.37)$$

for the initial condition which yields the expansion

$$X_k(\xi_1,\ldots,\xi_k,\lambda) = \sum_{n=0}^{\infty} \frac{\lambda^{2n}}{n!} X_{k,n}(\xi_1/\lambda,\ldots,\xi_k/\lambda) \qquad (3.38)$$

for the $k$-level correlation functions. The coefficient functions

$$X_{k,n}(\xi_1/\lambda,\ldots,\xi_k/\lambda) = \frac{(-1)^k}{\pi^k} \int G_k(s-\xi/\lambda, 1) w_{k,n}^{(0)}(s) B_k(s) d[s] \qquad (3.39)$$

depend only on the rescaled energies $\xi_p/\lambda$. The expansion (3.38) generalizes an observation made by Leyvraz and Seligman [10]. For a Poisson initial condition, they calculated the two level correlations in zeroth order by using a self-consistent perturbation theory and found that it depends only on $(\xi_1 - \xi_2)/\lambda$. Formula (3.38) shows that the coefficient functions for all indices $k$ of the correlation have such a scaling property provided the form of the expansion (3.37) is valid. An important caveat is in order: If the arbitrary ensemble has correlations, the zeroth order term of the expansion has to reproduce them. Since they are not on the scale $\xi/\lambda$, there should be an additional term to the zeroth order. However, this term will be specific for the ensemble under consideration and can therefore not be discussed in general. With the same caveat, the same scaling property can easily be derived for transitions toward the Gaussian Orthogonal or Symplectic Ensemble, GOE or GSE, respectively, again generalizing the corresponding observation made by Leyvraz and Seligman.

## 3.6 Translation Invariance and its Consequences

The correlation functions (3.35) are translation invariant as required provided the initial condition $z_k^{(0)}(s)$ depends only on the differences of the eigenvalues $s$. In order to show this, we shift the integration variables $s$ by the matrix $\xi$ in the integral (3.35). If $z_k^{(0)}(s)$ depends only on the differences of the eigenvalues, the correlations are obviously only functions of the differences $\xi_p - \xi_q$. This is exactly the translation invariance. Of course, the generating function $z_k^{(0)}(s)$ will depend on the differences of the eigenvalues if the the arbitrary correlations $\widehat{X}_k^{(0)}(\xi, \ldots, \xi_k)$ are translation invariant. Thus, the correlations of the transition are translation invariant if the arbitrary initial condition is translation invariant.

The fact that translation invariance is visible in the integral representation (3.35) has some helpful consequences. For the level density, it is shown in Appendix C that Eq. (3.35) directly implies

$$X_1(\xi, \lambda) = 1 \qquad (3.40)$$

for all values of $\lambda$. If we start from a properly unfolded arbitrary spectrum, the level density will always be unity during the entire transition, as it should be.

In the case of the two level correlations, translation invariance allows to reduce the four fold integral (3.35) to a double integral. We introduce the sum $\widetilde{r} = \xi_1 + \xi_2$ and the difference $r = \xi_1 - \xi_2$ of the energies. According to this rotation in energy space, we make the change

$$\begin{aligned} s_1 &= s_{11} + s_{21}, & t_1 &= s_{11} - s_{21}, \\ s_2 &= s_{12} + s_{22}, & t_2 &= s_{12} - s_{22} \end{aligned} \qquad (3.41)$$

of the integration variables. Since $z_2^{(0)}(s)$ depends only on the differences of the eigenvalues, it cannot depend on $s_1 + is_2$. We write $z_2^{(0)}(s) = z_2^{(0)}(s_1 - is_2, t_1, t_2)$ which, for reasons of consistency, should be even in each of the differences $t_1$ and $t_2$. As shown in Appendix D, the integrals over $s_1$ and $s_2$ can then be performed and the dependence on $\tilde{r}$ disappears. The two level correlations can thus be cast into the form

$$X_2(r,\lambda) = \frac{8}{\pi^3 \lambda^2} \int\limits_{-\infty}^{+\infty}\int\limits_{-\infty}^{+\infty} \exp\left(-\frac{1}{2\lambda^2}(t_1^2 + t_2^2)\right) \sinh\frac{rt_1}{\lambda^2} \sin\frac{rt_2}{\lambda^2}$$

$$\frac{t_1 t_2}{(t_1^2 + t_2^2)^2} \Im z_2^{(0)}(0, t_1, t_2) \, dt_1 dt_2 \qquad (3.42)$$

which is usually not amenable to further analytical treatment. For the higher correlations with $k > 2$, similar simplifications are likely to exist.

The expansion in the transition parameter discussed in the previous subsection can be written more explicitly for the two level correlations using the translation invariant form (3.42). Introducing $t_1/\lambda$ and $t_2/\lambda$ as new integration variables we find

$$X_2(r,\lambda) = \frac{8}{\pi^3} \int\limits_{-\infty}^{+\infty}\int\limits_{-\infty}^{+\infty} \exp\left(-\frac{1}{2}(t_1^2 + t_2^2)\right) \sinh\frac{r}{\lambda}t_1 \sin\frac{r}{\lambda}t_2$$

$$\frac{t_1 t_2}{(t_1^2 + t_2^2)^2} \frac{\Im z_2^{(0)}(0, \lambda t_1, \lambda t_2)}{\lambda^2} \, dt_1 dt_2 \, . \qquad (3.43)$$

We hold the rescaled energy difference $r/\lambda$ fixed and expand the initial condition in the transition parameter,

$$\frac{\Im z_2^{(0)}(0, \lambda t_1, \lambda t_2)}{\lambda^2} = \sum_{n=0}^{\infty} \frac{\lambda^{2n}}{(n+1)!} \left.\frac{\partial^{n+1} \Im z_2^{(0)}(0, \lambda t_1, \lambda t_2)}{\partial(\lambda^2)^{n+1}}\right|_{\lambda^2=0} \qquad (3.44)$$

which gives inserted into Eq. (3.43) the expansion

$$X_2(r,\lambda) = \sum_{n=0}^{\infty} \frac{\lambda^{2n}}{n!} X_{2,n}(r/\lambda) \qquad (3.45)$$

for the two level correlations. The coefficient functions

$$X_{2,n}(r/\lambda) = \frac{8}{\pi^3(n+1)} \int\limits_{-\infty}^{+\infty}\int\limits_{-\infty}^{+\infty} \exp\left(-\frac{1}{2}(t_1^2 + t_2^2)\right) \sinh\frac{r}{\lambda}t_1 \sin\frac{r}{\lambda}t_2$$

$$\frac{t_1 t_2}{(t_1^2 + t_2^2)^2} \left.\frac{\partial^{n+1} \Im z_2^{(0)}(0, \lambda t_1, \lambda t_2)}{\partial(\lambda^2)^{n+1}}\right|_{\lambda^2=0} dt_1 dt_2 \quad (3.46)$$

depend only on one variable, the rescaled energy difference $r/\lambda$. Of course, the caveat at the end of the last subsection is also valid here.

### 3.7 Level Number Variance

The variance $\Sigma^2$ of the level number on the unfolded energy scale studied as a function of the length $L$ of the energy interval under consideration is of great interest for the analysis of data since it gives information about the long range spectral correlation. It describes two level correlations and can therefore be expressed involving solely the two level cluster function [2]. This is also true for transitions and we can thus make use of the formula

$$\Sigma^2(L,\lambda) \;=\; L - 2\int_0^L (L-r)\, Y_2(r,\lambda)\, dr \;=\; -L(L-1) + 2\int_0^L (L-r)\, X_2(r,\lambda)\, dr \quad (3.47)$$

where we employed the relation $X_2(r,\lambda) = 1 - Y_2(r,\lambda)$ between the two level correlation and cluster function. Using Eq. (3.42) we can write down an integral representation for the level number variance for the transition from arbitrary correlations to GUE correlations. Since the $r$ integration can be carried out trivially, we arrive at the double integral

$$\begin{aligned}
\Sigma^2(L,\lambda) \;=\; & -L(L-1) + \frac{16\lambda^2}{\pi^3}\int_{-\infty}^{+\infty}\!\!\int_{-\infty}^{+\infty} \exp\left(-\frac{1}{2\lambda^2}(t_1^2+t_2^2)\right) \frac{t_1 t_2}{(t_1^2+t_2^2)^3} \\
& \left(\sinh\frac{L t_1}{\lambda^2}\sin\frac{L t_2}{\lambda^2} + \frac{2 t_1 t_2}{t_1^2+t_2^2}\left(1 - \cosh\frac{L t_1}{\lambda^2}\cos\frac{L t_2}{\lambda^2}\right)\right) \\
& \Im z_2^{(0)}(0,t_1,t_2)\, dt_1 dt_2 \qquad (3.48)
\end{aligned}$$

for all values of the transition parameter.

## 4 Transition as Diffusion

In Section 4.1, a general diffusion equation for the GUE is derived. The generalization to the GOE and the GSE is performed in Section 4.2. From these diffusive processes, stationary equations for the pure ensembles are constructed in Section 4.3. In Section 4.4 these equations are formulated on the unfolded energy scale.

### 4.1 Diffusion Equation

In the Eqs. (3.6) and (3.16), the generating function $Z_k(x+J,\alpha)$ is expressed as an integral over the initial condition $Z_k^{(0)}(s)$ and a Gaussian kernel in a Cartesean and in a curved space, respectively. This representation of the generating function allows to interpret the transition from arbitrary toward GUE correlations as a diffusive process. The squared transition parameter $\alpha^2$, i.e. the variance of the Gaussian distribution provides a fictitious time

$$t = \alpha^2/2 \qquad (4.1)$$

that controls the diffusion. The combination $x + J$ of the energies and the source variables plays the role of fictitious coordinates

$$r = x + J \qquad (4.2)$$

that define the space in which the diffusion takes place. We view these coordinates as the eigenvalues of a Hermitean supermatrix $\sigma_{\mathrm{fc}}$ which is diagonalized by a unitary supermatrix $u_{\mathrm{fc}}$ such that $\sigma_{\mathrm{fc}} = u_{\mathrm{fc}}^{-1} r u_{\mathrm{fc}}$. There is a gradient $\partial/\partial\sigma_{\mathrm{fc}}$ and a Cartesean Laplace operator

$$\Delta = \mathrm{trg}\frac{\partial^2}{\partial\sigma_{\mathrm{fc}}^2} \qquad (4.3)$$

in this space [18, 35, 36]. The graded probability density function (3.3) written in the form

$$Q_k(\sigma_{\mathrm{fc}}, t) = 2^{k(k-1)} \exp\left(-\frac{1}{2t}\mathrm{trg}\sigma_{\mathrm{fc}}^2\right) \qquad (4.4)$$

satisfies the definition of a diffusion kernel,

$$\frac{1}{2}\Delta Q_k(\sigma_{\mathrm{fc}}, t) = \frac{\partial}{\partial t} Q_k(\sigma_{\mathrm{fc}}, t) \qquad \text{and} \qquad \lim_{t \to 0} Q_k(\sigma_{\mathrm{fc}}, t) = \delta(\sigma_{\mathrm{fc}}) \ . \qquad (4.5)$$

Therefore, the integral (3.6) is equivalent to the diffusion equation

$$\frac{1}{2}\Delta Z_k(r, t) = \frac{\partial}{\partial t} Z_k(r, t) \qquad \text{with} \qquad \lim_{t \to 0} Z_k(r, t) = Z_k^{(0)}(r) \ . \qquad (4.6)$$

Since $Z_k(r, t)$ depends only on the eigenvalues $r$, the diffusion takes only place in this curved space and the radial part [18] of the Laplace operator is sufficient. It can be written as

$$\Delta_r = \frac{1}{B_k^2(r)} \frac{\partial}{\partial \vec{r}} \cdot B_k^2(r) \frac{\partial}{\partial \vec{r}} \qquad (4.7)$$

where we introduced the vector $\vec{r} = (r_{11}, \cdots, r_{k1}, r_{12}, \cdots, r_{k2})$ and the corresponding gradient $\partial/\partial\vec{r}$ for notational purposes. Notice that we choose the metric such that there are no imaginary units in front of the variables $r_{p2}$ and that the dot indicates the standard scalar product. Hence, we arrive at the diffusive process

$$\frac{1}{2}\Delta_r Z_k(r, t) = \frac{\partial}{\partial t} Z_k(r, t) \qquad \text{with} \qquad \lim_{t \to 0} Z_k(r, t) = Z_k^{(0)}(r) \qquad (4.8)$$

for the generating function of the transition correlations where the generating function of the arbitrary correlations is the initial condition. The level number $N$ does not appear in the diffusion equation itself, it is only implicitly contained in the initial condition. This means that asymptotic solutions for the generating function of the transition correlations can be found if the large $N$ behavior of the initial condition is known.

The diffusion is a physically very instructive picture. A Gaussian kernel propagates arbitrary correlations diffusively toward chaotic correlations. This formulation is completely equivalent to both integral representations, the original Cartesean (3.6) and the one in the curved eigenvalue space (3.16). The knowledge of the formula (3.11) is not necessary for the derivation of the diffusion equation (4.8). In fact, formula (3.11) was derived, in the ordinary [21] and in the supersymmetric case [18], by using a diffusion equation of the type (4.8). There, however, the fictitious time had no physical meaning, here, it is the transition parameter. The fictitious coordinates (4.2) combine energies and source variables, the relevant space is therefore not just given by the $k$ energies. The $k$ source variables are necessary for a construction of all together $2k$ independent variables $r$ which is essential for the procedure. The correlations functions (2.16), however, depend only on the energies. Hence, there is no diffusion of the type (4.8) for the correlation functions themselves, only for their generating functions. The existence of the diffusion process in the space of the fictitious coordinates (4.2) proves once more how well suited supersymmetry is in the theory of random matrices.

The special form of the Laplacean (4.7) allows the reformulation [18] of the diffusion in the curved space in a Cartesean space. A direct calculation shows that the function

$$W_k(r,t) = B_k(r) Z_k(r,t) \quad \text{with} \quad W_k^{(0)}(r) = B_k(r) Z_k^{(0)}(r) \tag{4.9}$$

as the initial condition solves the diffusion problem

$$\frac{1}{2}\frac{\partial^2}{\partial \vec{r}^2} W_k(r,t) = \frac{\partial}{\partial t} W_k(r,t) \quad \text{with} \quad \lim_{t \to 0} W_k(r,t) = W_k^{(0)}(r) . \tag{4.10}$$

The processes (4.8) and (4.10) have the formal solutions

$$Z_k(r,t) = \exp\left(\frac{t}{2}\Delta_r\right) Z_k^{(0)}(r)$$

$$W_k(r,t) = \exp\left(\frac{t}{2}\frac{\partial^2}{\partial \vec{r}^2}\right) W_k^{(0)}(r) \tag{4.11}$$

which could serve as a starting point for a power series expansion in the fictitious time $t$ if the integral representation (3.16) was unknown.

A comment on the boundary contribution in formula (3.16) is in order. One might question that the distribution $1 - \eta(r)$ satisfies Eq. (4.8). However, since $Z_k(r,t)$ in the form (3.6) and the integral in the curved space representation (3.16) solve the diffusion equation (4.8), $1 - \eta(r)$ must solve it as well. This can also be shown in a direct calculation using the integral representation of $\eta(r)$ which was constructed in Ref. [18], one easily finds $\Delta_r \eta(r) = 0$ as it should be since $\eta(r)$ is not time dependent. In this sense, the boundary contribution $1 - \eta(r)$ which ensures the normalization of the generating function is a stationary point of the diffusive process. Nevertheless, a minor problem arises when the transition parameter vanishes. As discussed already in Section 3.2, the boundary contribution is obsolete at $t = 0$. A careful investigation of the power series in $t$ resulting from the formal solution (4.11) proves that this new type of singularity is an artifact caused by the incomplete treatment of the boundary contributions of the real parts of the Green functions. We therefore do not worry about this further since we are only interested in the imaginary parts and the limit of infinitely many levels which are both known to come out correct [25]. It should be stressed that Rothstein [22] showed a cleaner way of treating the boundary contributions by modifying the integration measure in the curved space such that it still has the symmetries of the original, in our case Cartesean, space. Unfortunately, so far it was only possible to apply this technique [25] to the one point function, i.e. to $2 \times 2$ Hermitean supermatrices.

## 4.2 Orthogonal and Symplectic Case

Besides the initial condition, the only input for the derivation of the diffusion equation is the observation that the Gaussian probability density (2.9) is mapped onto a graded probability density (3.3) which is as well Gaussian. In Cartesean coordinates, this feature is also present [16, 17] in the Gaussian Orthogonal (GOE) and in the Gaussian Symplectic Ensemble (GSE). Hence there exists a Cartesean integral representation of the type (3.6) for the generating functions

$$Z_{\beta k}(x + J, t) = \int Q_{\beta k}(\sigma - x - J, t) Z_{\beta k}^{(0)}(s) d[\sigma] \qquad (4.12)$$

of the correlations in the case of transitions toward all three of the Gaussian Ensembles. The parameter $\beta$ takes the values 1, 2 or 4 for the orthogonal, unitary or symplectic symmetry class, respectively. Here and in the following two subsections, all probability density functions and all correlation functions and their generating functions have the additional index $\beta$ which indicates a proper adjustment of the definitions in Section 2 to the different symmetry classes. The variance of the Gaussian distribution for the three values of $\beta$ is often written as $2v^2/\beta$ with Eq. (2.9) as special case for $\beta = 2$. Thus, fixing the energy scale through the condition $\sqrt{2}v = 1$ yields an effective variance of $1/\beta$. For all three symmetry classes we use the definition (4.1) of the fictitious time $t$. Hence, the variance of the graded probability density function reads $2t/\beta$. For the GOE and the GSE, the supermatrices $\sigma$ have to have dimension $4k \times 4k$ in order to implement the additional symmetry constraints. The diagonalization analogous to (3.7) involves properly defined $4k \times 4k$ matrices $u$ and the eigenvalue matrices

$$\begin{aligned}
s &= \operatorname{diag}(s_{111}, s_{121}, is_{12}, is_{12}, \cdots, s_{k11}, s_{k21}, is_{k2}, is_{k2}) && \text{(GOE)} \\
s &= \operatorname{diag}(s_{11}, s_{11}, is_{112}, is_{122}, \cdots, s_{k1}, s_{k1}, is_{k12}, is_{k22}) && \text{(GSE)}
\end{aligned} \quad (4.13)$$

with $3k$ independent variables. However, the $4k \times 4k$ diagonal matrices of the energies and the source variables

$$\begin{aligned}
x &= \operatorname{diag}(x_1, x_1, x_1, x_1, \cdots, x_k, x_k, x_k, x_k) \\
J &= \operatorname{diag}(-J_1, -J_1, +J_1, +J_1, \cdots, -J_k, -J_k, +J_k, +J_k)
\end{aligned} \quad (4.14)$$

contain only $k$ independent variables each. The combination $x + J$ has only $2k$ independent variables and cannot serve as the fictitious coordinate space. We have to make the replacement

$$x + J \longrightarrow r \quad (4.15)$$

where $r$ has the same form as $s$ in Eq. (4.13). For notational purposes, we introduce the vectors

$$\begin{aligned}
\vec{r} &= (r_{111}, r_{121}, r_{12}, \cdots, r_{k11}, r_{k21}, r_{k2}) && \text{(GOE)} \\
\vec{r} &= (r_{11}, r_{112}, r_{122}, \cdots, r_{k1}, r_{k12}, r_{k22}) && \text{(GSE)}
\end{aligned} \quad (4.16)$$

and the corresponding gradients. Moreover, by writing $\sigma_{\text{fc}} = u_{\text{fc}}^{-1} r u_{\text{fc}}$ with $\sigma_{\text{fc}}$ and $u_{\text{fc}}$ of the same form as $\sigma$ and $u$, respectively, we may define a Cartesean Laplace operator $\Delta_\beta$ analogously to Eq. (4.3). Obviously, the diffusion takes only place in the curved space of the eigenvalues $r$ and the radial part

$$\Delta_{\beta r} = \frac{1}{B_{\beta k}(r)} \frac{\partial}{\partial \vec{r}} \cdot B_{\beta k}(r) \frac{\partial}{\partial \vec{r}} \quad (4.17)$$

suffices. The Berezinians of the transformation to eigenvalue angle coordinates can be worked out as in Ref. [18],

$$\begin{aligned}
B_{1k}(r) &= \frac{\prod_{p=1, j>l}^{k} |r_{pj1} - r_{pl1}| \prod_{p>q, j, l} |r_{pj1} - r_{ql1}| \prod_{p>q} (ir_{p2} - ir_{q2})^4}{\prod_{p, q, j} (r_{pj1} - ir_{q2})^2} \\
B_{4k}(r) &= \frac{\prod_{p>q} (r_{p1} - r_{q1})^4 \prod_{p=1, j>l}^{k} |ir_{pj2} - ir_{pl2}| \prod_{p>q, j, l} |ir_{pj2} - ir_{ql2}|}{\prod_{p, q, j} (r_{p1} - ir_{qj2})^2}
\end{aligned} \quad (4.18)$$

for the GOE and the GSE and $B_{2k}(r) = B_k^2(r)$ for the GUE. Thus, collecting everything we arrive at the diffusion equation

$$\frac{\beta}{4} \Delta_{\beta r} Z_{\beta k}(r, t) = \frac{\partial}{\partial t} Z_{\beta k}(r, t) \quad \text{with} \quad \lim_{t \to 0} Z_{\beta k}(r, t) = Z_{\beta k}^{(0)}(r) \quad (4.19)$$

for all three symmetry classes and for arbitrary initial conditions. In case of the GOE and the GSE, the fictitious coordinates $r$ define according to Eq. (4.15) an extended space in which the generating function is propagated. In order to calculate the correlation functions, we have to restrict this space to the subspace defined by the combination $x + J$. The formal solution of the diffusion equation

$$Z_{\beta k}(r, t) = \exp\left(\frac{\beta t}{4}\Delta_{\beta r}\right) Z_{\beta k}^{(0)}(r) \qquad (4.20)$$

allows an expansion in powers of the fictitious time $t$ for all three values of $\beta$. In order to calculate the correlation functions up to $n$-th order in $t$, i.e. in $\alpha^2$, one has to evaluate $\Delta_{\beta r}^{n'} Z_{\beta k}^{(0)}(r), n' = 0, \ldots, n$ at $x + J$.

It should be emphasized that the diffusion equation (4.19) was derived without performing the angular average, i.e. without knowing the explicit solution of the supersymmetric Harish-Chandra-Itzykson-Zuber integral

$$\Gamma_{\beta k}(s, r, t) = \int Q_{\beta k}(u^{-1}su - v^{-1}rv, t) \, d\mu(u) \qquad (4.21)$$

in the cases $\beta = 1$ and $\beta = 4$. Unfortunately, the explicit solution is only known in the unitary case (3.11). Even for the ordinary orthogonal and symplectic groups, this integral has not been evaluated yet. In other words, we cannot give an eigenvalue integral representation of the type (3.16) for transitions toward the GOE and the GSE. So far, the solution (4.20) is the only available one. This is closely related to the structure of the radial part (4.7) of the Laplacean. For $\beta = 1$ and $\beta = 4$, there is no ansatz in the spirit of Eq. (4.9) that would reduce the diffusion in the curved space (4.19) to a diffusion in a Cartesean space analogously to Eq. (4.10). As mentioned already, the explicit solution (3.11) of the group integral was derived [18] by using this property of the radial part in the unitary case.

However, for the sake of completeness, we summarize the general formulae involving the kernel (4.21). By construction, it has to satisfy the diffusion equation

$$\frac{\beta}{4}\Delta_{\beta r}\Gamma_{\beta k}(s, r, t) = \frac{\partial}{\partial t}\Gamma_{\beta k}(s, r, t) \qquad (4.22)$$

with the initial condition

$$\lim_{t \to 0} \Gamma_{\beta k}(s, r, t) = \int \delta(u^{-1}su - v^{-1}rv) \, d\mu(u)$$

$$= \frac{\det[\delta(s_{p1} - r_{q1})]_{p,q=1,\ldots,k} \det[\delta(s_{p2} - r_{q2})]_{p,q=1,\ldots,k}}{\sqrt{B_{\beta k}(s)B_{\beta k}(r)}} \qquad (4.23)$$

and, moreover, it has the property

$$\Gamma_{\beta k}(r, r', t + t') = \int \Gamma_{\beta k}(r, s, t) \Gamma_{\beta k}(s, r', t') B_{\beta k}(s) d[s] . \qquad (4.24)$$

Thus, we can write the solution of Eq. (4.19) in the form

$$Z_{\beta k}(r,t) = \int \Gamma_{\beta k}(s,r,t) Z^{(0)}_{\beta k}(s) B_{\beta k}(s) d[s] \qquad (4.25)$$

which of course coincides with Eq. (3.16) in the case $\beta = 2$.

## 4.3 Stationary Equations for the Pure Ensembles

The diffusion equation (4.19) allows for arbitrary initial conditions. As a special case, we can also study the diffusion of the generating function of a symmetry class into itself. More precise, for a given $\beta$, we take the generating function of the correlations of this symmetry class as the initial condition of the Eq. (4.19),

$$Z^{(0)}_{\beta k}(r) = Z^{(1)}_{\beta k}(r) \qquad (4.26)$$

where, as in Section 2.2, the upper index (1) indicates the pure Gaussian Ensemble. This situation is somewhat stationary since the propagation changes neither the symmetry nor the structure of the correlation. Of course, the energy scale is affected and becomes a function of the fictitious time.

It is very easy to determine this rescaling from the definitions (2.16), (2.17) and (2.19) of the correlation and the generating function without using the diffusion equation or any explicit form of the correlation functions. Both probability density functions $P^{(j)}_{\beta N}(H^{(j)})$ $j = 0, 1$ are now Gaussian. Again, we move the transition parameter $\alpha$ into the standard deviation of $P^{(1)}_{\beta N}(H^{(1)})$. By defining $H^{(1)\prime} = H^{(0)} + H^{(1)}$ as new matrices for the integration, the $H^{(0)}$ integral becomes the convolution of the two Gaussians with variances $2v^2/\beta$ and $2v^2\alpha^2/\beta$ which gives a Gaussian with variance $2v^2(1+\alpha^2)/\beta$. By using the elements of the matrices $H^{(1)}/\sqrt{1+\alpha^2}$ as new integration variables, we see that the energies and source variables have to be rescaled by $\sqrt{1+\alpha^2}$. As always, we set $\sqrt{2}v = 1$. Thus, in a given symmetry class, we find the scaling law

$$\widehat{R}_{\beta k}(x_1, \ldots, x_k, \alpha) = \frac{1}{\sqrt{1+\alpha^2}^k} \widehat{R}^{(1)}_{\beta k}\left(\frac{x_1}{\sqrt{1+\alpha^2}}, \ldots, \frac{x_k}{\sqrt{1+\alpha^2}}\right) \qquad (4.27)$$

for the transition of the correlations into themselves. It follows just from the definition of the correlation functions. The knowledge of the explicit solutions [3] for the pure ensembles is not necessary to derive Eq. (4.27). Also only employing the definition, we arrive at the scaling law

$$Z_{\beta k}(r,t) = Z^{(1)}_{\beta k}\left(\frac{r}{\sqrt{1+2t}}\right) \qquad (4.28)$$

for the generating functions. Here, we use the fictitious time $t = \alpha^2/2$ and the fictitious coordinates $r$.

We now take advantage of the fact that the generating function satisfies the diffusion equation (4.19). For our special choice of the initial conditions, we have

$$\frac{\beta}{4}\Delta_{\beta r} Z^{(1)}_{\beta k}\left(\frac{r}{\sqrt{1+2t}}\right) = \frac{\partial}{\partial t} Z^{(1)}_{\beta k}\left(\frac{r}{\sqrt{1+2t}}\right) \qquad (4.29)$$

with the trivially fulfilled initial condition

$$\lim_{t\to 0} Z^{(1)}_{\beta k}\left(\frac{r}{\sqrt{1+2t}}\right) = Z^{(1)}_{\beta k}(r) . \qquad (4.30)$$

We set $r_t = r/\sqrt{1+2t}$ and $\vec{r}_t = \vec{r}/\sqrt{1+2t}$ and rewrite the time derivative

$$\frac{\partial}{\partial t} Z^{(1)}_{\beta k}(r_t) = \frac{\partial Z^{(1)}_{\beta k}(r_t)}{\partial \vec{r}_t} \cdot \frac{\partial \vec{r}_t}{\partial t} = -\frac{\vec{r}_t}{1+2t} \cdot \frac{\partial Z^{(1)}_{\beta k}(r_t)}{\partial \vec{r}_t} . \qquad (4.31)$$

Since the Laplacean obeys $\Delta_{\beta r} = \Delta_{\beta r_t}/(1+2t)$ we find

$$\frac{\beta}{4}\Delta_{\beta r_t} Z^{(1)}_{\beta k}(r_t) = -\vec{r}_t \cdot \frac{\partial}{\partial \vec{r}_t} Z^{(1)}_{\beta k}(r_t) \qquad (4.32)$$

which contains no explicit time dependence anymore. Hence, the generating function satisfies a stationary partial differential equation in the rescaled fictitious coordinates $r_t$. Therefore, we drop the index $t$ and

$$\left(\frac{\beta}{4}\Delta_{\beta r} + \vec{r}\cdot\frac{\partial}{\partial \vec{r}}\right) Z^{(1)}_{\beta k}(r) = 0 \qquad (4.33)$$

is our final result. The boundary condition is of course the normalization $Z^{(1)}_{\beta k}(x) = 1$. This stationary equation is equivalent to the original integral definition of the type (2.7) for the pure ensembles. Notice that the only input is the definition of the correlation functions and the existence of the diffusion equation (4.19) for arbitrary initial conditions. It is easy to rewrite the stationary equation as

$$\left(\frac{\beta}{4}\frac{\partial}{\partial \vec{r}} + \vec{r}\right)\cdot B_{\beta k}(r)\frac{\partial Z^{(1)}_{\beta k}(r)}{\partial \vec{r}} = 0 \qquad (4.34)$$

which has an integrating factor such that

$$\frac{\partial}{\partial \vec{r}} \cdot \exp\left(+\frac{\beta}{2}\mathrm{tr} r^2\right) B_{\beta k}(r)\frac{\partial Z^{(1)}_{\beta k}(r)}{\partial \vec{r}} = 0 . \qquad (4.35)$$

From this equation, a new integral representation for $Z^{(1)}_{\beta k}(r)$ can be constructed. For a deeper understanding of the stationary equation, it is helpful to discuss it explicitly. In Appendix B, we study the simplest case, the GUE, as an example.

## 4.4 Diffusion Equation on the Unfolded Energy Scale

From the previous discussion and from Section 3.4, it is obvious that the transition on the unfolded energy is a diffusive process as well. Naturally, the fictitious time is now given by

$$\tau = \lambda^2/2 \tag{4.36}$$

and the combination of rescaled energies $\xi$ and source variables $j$ provides the fictitious coordinate space. We write

$$\rho = \xi + j \tag{4.37}$$

with $\rho$ having the form (3.8) in the case of the GUE and we make the replacement

$$\xi + j \longrightarrow \rho \tag{4.38}$$

where $\rho$ is of the form (4.13) in the case of the GOE and GSE, respectively. The structure of the curved space in which the diffusion takes place remains unchanged and so does the Laplace operator, we simply have to replace $r$ with $\rho$ in Eq. (4.17). Hence, the generating functions

$$\begin{aligned} z_{\beta k}^{(0)}(\rho) &= \lim_{N \to \infty} Z_{\beta k}^{(0)}(r) \\ z_{\beta k}(\rho, \tau) &= \lim_{N \to \infty} Z_{\beta k}(r, t) \end{aligned} \tag{4.39}$$

obey the diffusion equation

$$\frac{\beta}{4} \Delta_{\beta \rho} \, z_{\beta k}(\rho, \tau) = \frac{\partial}{\partial \tau} z_{\beta k}(\rho, \tau) \qquad \text{with} \qquad \lim_{\tau \to 0} z_{\beta k}(\rho, \tau) = z_{\beta k}^{(0)}(\rho) \tag{4.40}$$

for all three symmetry classes and for arbitrary initial conditions. The formal solution

$$z_{\beta k}(\rho, \tau) = \exp\left(\frac{\beta \tau}{4} \Delta_{\beta \rho}\right) z_{\beta k}^{(0)}(\rho) \tag{4.41}$$

is completely analogous to Eq. (4.20). In the case of the GUE, a separation of the type (4.9) is of course also possible. The diffusive processes on the original and the unfolded scale are the same, the initial conditions are different. This is no surprise since the Laplacean (4.17) is independent of the level number $N$ which appears only in the initial condition. Once more, the space of the eigenvalues of the supermatrices proves the appropriate one since it describes just the correlations and does not care about the scale on which they are measured. In other words, a once unfolded initial condition remains always unfolded, even during and after the propagation into the regime dominated by the Gaussian ensembles. It should be mentioned that the diffusion equation on the unfolded scale is more general than the one on the original scale. This is because $\lambda$, and thus $\tau$, as discussed in Section 3.4, turn out to be the relevant parameter on the unfolded scale for a wide class of parameterizations (2.24) on the original scale.

Since the diffusion equations on the original and on the unfolded scale are the same, the integral representation (4.25) has to hold on the unfolded scale as well,

$$z_{\beta k}(\rho, \tau) = \int \Gamma_{\beta k}(s, \rho, \tau) z_{\beta k}^{(0)}(s) B_{\beta k}(s) d[s] \qquad (4.42)$$

which becomes Eq. (3.32) in the case $\beta = 2$.

The stationary equations for the pure ensembles on the unfolded scale can directly be read of from the diffusion (4.40). In the limit $\tau \to \infty$, the correlations are purely chaotic and thus independent of $\tau$. The same must be true for the generating function,

$$\lim_{\tau \to \infty} z_{\beta k}(\rho, \tau) = z_{\beta k}^{(1)}(\rho) = \lim_{N \to \infty} Z_{\beta k}^{(1)}(r) . \qquad (4.43)$$

Consequently, the derivative with respect to $\tau$ in Eq. (4.40) vanishes and we conclude

$$\Delta_{\beta \rho} z_{\beta k}^{(1)}(\rho) = 0 . \qquad (4.44)$$

The boundary conditions are such that the correlations are translation invariant, this implies that the generating function can depend only on the differences of the eigenvalues $\rho$.

By using Dyson's Brownian motion picture [27], hierarchic relations were constructed in Ref. [9] for the correlation functions on the original and the unfolded scale, $R_k(x_1, \ldots, x_k, t)$ and $X_k(\xi_1, \ldots, \xi_k, \tau)$, respectively, in which the energies provide the fictitious coordinates. These equations couple the correlations of the order $k$ to those of the order $k + 1$ and thus cannot be viewed as a diffusion. Due to the application of supersymmetry and the graded eigenvalue method, our equations for the generating functions describe an independent diffusion for every order $k$ in the extended fictitious coordinate space.

## 5 Transition from Poisson Regularity to Chaos

In Section 5.1, we discuss some general aspects of regularity in random matrix theory. The Poisson Ensemble is introduced in Section 5.2. The generating function is worked out in Section 5.3. In Section 5.4, the correlation functions are evaluated. In section 5.5, the two level correlation function is discussed in detail.

### 5.1 First Considerations

The Gaussian Ensembles model quantum systems whose eigenenergies repel each other because of some interaction. This famous Wigner repulsion, now often referred to as quantum chaos, manifests itself in the simple fact that the matrices $H^{(1)}$ which form the Gaussian Ensembles are fully occupied. Since no basis is preferred, the GOE, GUE and GSE are invariant under $O(N)$, $U(N)$ and $Sp(N)$ rotations, respectively. In other words, the probability density functions depend only on the eigenvalues. As well known [3], the non-trivial part of the ensemble average becomes then an integration over the eigenvalues $x_n^{(1)}$, $n = 1, \ldots, N$ involving the Jacobian $|\Delta_N(x^{(1)})|^\beta$. The Vandermonde determinant $\Delta_N(x^{(1)}) = \prod_{n>m}(x_n^{(1)} - x_m^{(1)})$ annihilates the whole integrand whenever two eigenenergies degenerate. This is the Wigner repulsion. The chaotic features of the correlations on the unfolded energy scale are almost solely due to this effect. The special form of the probability density function is not important as long as it is rotation invariant [3]. This robustness or universality of the chaotic fluctuation properties has recently been verified in detailed studies for spectral correlations [28] and also for scattering systems [29].

How to construct a random matrix model for quantum systems which do not have chaotic, but regular fluctuation properties? This question has been studied by numerous authors addressing many different physical situations, especially those which show a transition [9] toward chaotic fluctuation properties. As already stated in the introduction, in this work, we are mainly interested in a regular system that does not show any spectral correlations. The Poisson Ensemble models such a system. In view of the discussion above, it comes as no surprise that ensembles without spectral correlations are efficiently designed by breaking the rotation invariance. Loosely speaking, the influence of the Vandermonde determinant becomes the weaker the stronger the rotation invariance is broken. This models the vanishing of the Wigner repulsion. A transition ensemble of particular interest is the Band Random Matrix Ensemble in which all entries of the matrices beyond a certain distance from the diagonal are set equal to zero. For a fixed, large dimension of the matrices, the fluctuations become more and more Poisson regular as the bandwidth is made smaller [30, 31]. Closely related and maybe better suited in a wide class of physical situations are Gaussian Ensembles which are coupled in such a way that strength is transported between next neighbors. Those ensembles consist of tridiagonal matrices whose entries are again random matrices drawn from Gaussian Ensembles. Here, for a fixed dimension, the ratio between the variances of the Gaussian probability density functions for the side and the main diagonal controls the transition. These ensembles are being used as models for systems which show localization effects and to study the fluctuation properties of mesoscopic particles [32, 33]. Recently, an alternative ensemble was studied [34] that is in a certain sense rotation invariant but still yields an interpolation between Poisson regular and chaotic fluctuation properties. Here, however, we are interested in transition ensembles of the form (2.14). Hence, we have to construct an ensemble of matrices $H^{(0)}$ that models pure Poisson regular fluctuation properties.

## 5.2 Poisson Ensemble

The easiest way to break the rotation invariance is to assign a fixed value to certain matrix elements. If we want to eliminate all correlations we simply have to remove all interactions from the matrices $H^{(0)}$ by setting all offdiagonal matrix elements equal to zero,

$$P_N^{(0)}(H^{(0)}) = \prod_{n=1}^{N} p^{(0)}(H_{nn}^{(0)}) \prod_{n>m} \delta(\mathrm{Re}H_{nm}^{(0)})\delta(\mathrm{Im}H_{nm}^{(0)}) \qquad (5.1)$$

where $p^{(0)}(z)$ is a smooth, symmetric, but otherwise arbitrary, probability density function with the normalization

$$\int_{-\infty}^{+\infty} p^{(0)}(z)\,dz = 1 \;. \qquad (5.2)$$

This ensemble is referred to as Poisson Ensemble. The lack of any correlation is straightforwardly shown. From definition (2.2) we find

$$R_1^{(0)}(x) = N\,p^{(0)}(x) \qquad (5.3)$$

for the level density and

$$\begin{aligned} R_k^{(0)}(x_1,\ldots,x_k) &= \frac{N!}{(N-k)!} \prod_{p=1}^{k} p^{(0)}(x_p) = \frac{N!}{(N-k)!N^k} \prod_{p=1}^{k} R_1^{(0)}(x_p) \\ &= \prod_{p=1}^{k} \left(1 - \frac{p-1}{N}\right) R_1^{(0)}(x_p) \end{aligned} \qquad (5.4)$$

for the $k$-level correlation functions. As mentioned already in Section 2.1, we always assume $x_p \neq x_q$ for $p \neq q$ in order to remove the trivial contributions containing $\delta(x_p - x_q)$. Obviously, there are no true correlations, because the $k$-level correlation function is simply a $k$-fold product of level densities. Notice that the prefactors in Eq. (5.4) ensure the normalization of the $k$-level correlation function to $N!/(N-k)!$ which holds by definition for arbitrary ensembles. For the same reason, the prefactor in front of the product of all level densities is unity in the decomposition of the higher order Gaussian correlation functions into the lower order ones according to Eq. (2.11). It is therefore not very convenient to define $k$-level cluster functions in the Poisson case. On the unfolded energy scale, however, all these normalization factors become unity and we can write

$$\begin{aligned} X_k^{(0)}(\xi_1,\ldots,\xi_k) &= 1 \qquad \text{for all } k \\ Y_k^{(0)}(\xi_1,\ldots,\xi_k) &= 0 \qquad \text{for } k > 1 \end{aligned} \qquad (5.5)$$

without causing inconsistencies. These equations can also be viewed as the definition of a Poisson Ensemble. The inclusion of the real part of the Green function yields the contribution of a Cauchy principal value integral,

$$\widehat{R}_1^{(0)}(x) = \frac{N}{\pi} \int\limits_{-\infty}^{+\infty} P\left(\frac{1}{x-z}\right) p^{(0)}(z) dz \mp i N p^{(0)}(x) . \tag{5.6}$$

All other formulae above hold the same if we replace $R_1(x)$ with $\widehat{R}_1(x)$.

## 5.3 Generating Function

As the initial condition for the diffusion equation or the the integral representation (3.16) we need to evaluate the generating function (3.15) for the Poisson Ensemble (5.1). We find

$$Z_k^{(0)}(s) = \int P_N^{(0)}(H^{(0)}) \det{}^{-1} \left(s^\pm \otimes 1_N - 1_{2k} \otimes H^{(0)}\right) d[H^{(0)}]$$
$$= \left(\zeta_k^{(0)}(s)\right)^N \tag{5.7}$$

since all diagonal elements are equally distributed. There are various ways to compute the integral

$$\zeta_k^{(0)}(s) = \int\limits_{-\infty}^{+\infty} p^{(0)}(z) \prod_{p=1}^{k} \frac{i s_{p2} - z}{s_{p1}^\pm - z} dz , \tag{5.8}$$

the easiest one is probably to use the identity

$$\prod_{p=1}^{k} \frac{\nu_p - z}{\mu_p - z} = 1 + \sum_{p=1}^{k} \frac{\prod_{q=1}^{k}(\nu_q - \mu_p)}{\prod_{q \neq p}(\mu_q - \mu_p)} \frac{1}{\mu_p - z} \tag{5.9}$$

which holds for an arbitrary set of variables $\mu_p, \nu_p$, $p = 1, \ldots, k$. Degeneracies of the type $\mu_p = \nu_q$ for some $p, q$ are taken care of through cancelations in the right hand side. Even degeneracies of the type $\mu_p = \mu_q$ for some $p, q$ come out correct if viewed as the limit $\mu_p \to \mu_q$. The identity (5.9) can be proven by induction or by standard methods of complex analysis. Hence, for the integral (5.8) we find

$$\zeta_k^{(0)}(s) = 1 + \sum_{p=1}^{k} b_p(s) \int\limits_{-\infty}^{+\infty} \frac{p^{(0)}(z)}{s_{p1}^\pm - z} dz = 1 + \frac{\pi}{N} \sum_{p=1}^{k} b_p(s) \widehat{R}_1^{(0)}(s_{p1}) \tag{5.10}$$

where the function

$$b_p(s) = \frac{\prod_{q=1}^{k}(i s_{q2} - s_{p1})}{\prod_{q \neq p}(s_{q1}^\pm - s_{p1}^\pm)} = (i s_{p2} - s_{p1}) \prod_{q \neq p} \frac{i s_{q2} - s_{p1}}{s_{q1}^\pm - s_{p1}^\pm} \tag{5.11}$$

can be considered as the projection of $B_k(s)$, the square root of the Berezinian, onto the coordinate axis $s_{p1}$. It is interesting to remark that these functions obey the sum rule

$$\sum_{p=1}^{k} b_p(s) = -\mathrm{tr} gs \ . \tag{5.12}$$

Collecting everything we arrive at

$$Z_k^{(0)}(s) = \left(1 + \frac{\pi}{N}\sum_{p=1}^{k} b_p(s)\widehat{R}_1^{(0)}(s_{p1})\right)^N \ . \tag{5.13}$$

It is easily checked that $Z_k^{(0)}(x+J)$ is indeed the generating function for the Poisson Ensemble. We have $b_p(x+J) = 2J_p \widetilde{b}_p(x+J)$ with $\widetilde{b}_p(x) = 1$, the derivatives of $\widetilde{b}_p(x+J)$ cannot produce expressions proportional to $J_p$ or $1/J_p$. Hence, the only term that contributes to the correlation functions (2.6) is the one containing the product $\prod_{p=1}^{k} J_p$. Counting the prefactors correctly we find the results given in the last subsection.

Before we can compute the generating function (3.33) on the unfolded scale, we have to specify the one point function $\widehat{R}_1^{(0)}(x)$, in particular its level number dependence. As discussed in Section 2.2 the two level densities involved, $R_1^{(0)}(x)$ and $R_1^{(1)}(x)$, ought to have the same asymptotic $N$ behavior. This suggests to choose the GUE level density for the Poisson Ensemble. We set $R_1^{(0)}(x) = R_1^{(1)}(x)$ which implies through the Kramers Kronig relation also the equality of the real parts. For large $N$, we can use the Pastur equation of Section 3.3 to determine the one point function for the transition which is, as discussed, just given by Eq. (3.29). On the unfolded scale, we are only interested in the transition parameter $\alpha$ of the order $1/\sqrt{N}$. Consequently, the $\alpha$ dependence vanishes for large $N$ and we have

$$\widehat{R}_1(x,\alpha) \longrightarrow \widehat{R}_1^{(0)}(x) = \widehat{R}_1^{(1)}(x) \longrightarrow \frac{1}{\pi}\left(x \mp i\sqrt{2N-x^2}\right) \tag{5.14}$$

and thus $D = 1/R_1(0,\alpha) = \pi/\sqrt{2N}$. The equations (5.7) and (5.10) yield in formula (3.33),

$$z_k^{(0)}(s) = \exp\left(\lim_{N\to\infty} N \ln \zeta_k^{(0)}(Ds)\right) = \exp\left(\lim_{N\to\infty} N \ln\left(1 + i\frac{\pi}{N}\sum_{p=1}^{k} \mp b_p(s)\right)\right)$$

$$= \exp\left(i\pi \sum_{p=1}^{k} \mp b_p(s)\right) = \prod_{p=1}^{k} \exp\left(\mp i\pi b_p(s)\right) \tag{5.15}$$

where we used $b_p(Ds) = Db_p(s)$ and $\widehat{R}_1^{(0)}(Ds_{p1}) \to \mp i/D$ for $N \to \infty$. The signs are determined by the choice of the sign of the imaginary increment in the Green function.

## 5.4 Correlation Functions

In order to evaluate the $k$-level correlation functions, we insert the result (5.13) into Eq. (3.18) and arrive at

$$\widehat{R}_k(x_1,\ldots,x_k,\alpha) = \frac{(-1)^k}{\pi^k} \int G_k(s-x,\alpha) \left(1 + \frac{\pi}{N}\sum_{p=1}^{k} b_p(s)\widehat{R}_1^{(0)}(s_{p1})\right)^N B_k(s)d[s] \quad (5.16)$$

for non-zero $\alpha$. As pointed out in Section 3.2, the limit $\alpha \to 0$ yields the correct result for the generating function, i.e. here in this case the generating function (5.13). As a support of the general discussion in previous sections and as a further test of our calculation, it is worthwhile to check the opposite limit for the correlation functions (5.16). Following Section 2.2 we are free to replace the eigenvalues $s$ in the initial condition with $s/\sqrt{1-\alpha^2}$. Since $Z_k^{(0)}(s)$ is a continuous function of $\zeta_k^{(0)}(s)$, we only need to check the limit in the rescaled function $\zeta_k^{(0)}(s/\sqrt{1-\alpha^2})$. It is instructive to do that explicitly in the form (5.10). Because of the homogeneity of $b_p(s)$ the relevant limit reduces to

$$\lim_{\alpha \to 1} \frac{1}{\sqrt{1-\alpha^2}} \widehat{R}_1\left(\frac{s_{p1}}{\sqrt{1-\alpha^2}}\right) = \lim_{\alpha \to 1} \frac{N}{\pi} \int_{-\infty}^{+\infty} \frac{p^{(0)}(z)}{s_{p1}^{\pm} - \sqrt{1-\alpha^2}z} dz = \frac{N}{\pi} \frac{1}{s_{p1}^{\pm}} \quad (5.17)$$

and gives after resummation of the expansion in Eq. (5.10) the expression

$$\lim_{\alpha \to 1} Z_k^{(0)}\left(\frac{s}{\sqrt{1-\alpha^2}}\right) = \prod_{p=1}^{k} \left(\frac{is_{p2}}{s_{p1}^{\pm}}\right)^N \quad (5.18)$$

which is indeed the correct result [18] for the GUE. Notice that the limit (5.17) is just the unaveraged Green function (3.26) which has already been used in the discussion of the Pastur equation in Section 3.3.

Lenz [11] presented a calculation of the correlation functions for the Poisson GUE transition employing the Mehta Dyson method, i.e. without using supersymmetry. After several changes of variables, he found an integral representation for the one level correlations which coincides exactly with our formula (5.16) for $k = 1$. For the two level correlations, he constructed an integral representation which contains a ratio of level number dependent determinants. Unfortunately, this representation seems not to be amenable to further analytical treatment. In particular, it is not clear how to take the limit of infinetly many levels. Hence, the advantage of supersymmetry and the graded eigenvalue becomes obvious in the simple form of the generating function (5.13). The crucial quantities are the functions $b_p(s)$ which are related to the Berezinian. They generate the determinants in Lenz's integral representation and allow to take the limit $N \to \infty$ directly in the form (5.15).

The construction of the physically interesting correlations $R_k(x_1, \ldots, x_k, \alpha)$ is not completely straightforward. In order to apply the procedure described in Section 2.1, the contribution from the Poisson Ensemble in the integrand has to be expanded in a binomial series. No further simplifications seem possible in this procedure. Fortunately, things become simpler on the more important unfolded energy scale. For non-zero $\lambda$ we find from Eqs. (3.35) and (5.15)

$$\widehat{X}_k(\xi_1, \ldots, \xi_k, \lambda) = \frac{(-1)^k}{\pi^k} \int G_k(s - \xi, \lambda) \prod_{p=1}^{k} \exp(\mp i\pi b_p(s)) \, B_k(s) d[s] \quad (5.19)$$

which implies certain simplifications due to the product structure in the integrand. Using the symbol $\Im$, we can write

$$X_k(\xi_1, \ldots, \xi_k, \lambda) = \frac{(-1)^k}{\pi^k} \int G_k(s - \xi, \lambda) \Im \prod_{p=1}^{k} \exp(\mp i\pi b_p(s)) \, B_k(s) d[s] \quad (5.20)$$

representing the $k$-level correlation functions on the unfolded energy scale for the transition from Poisson regularity to chaos as a $2k$-fold integral. Since the generating function (5.15) depends only on the differences of the eigenvalues, the correlation functions are indeed translation invariant as discussed in Section 3.4.

In the case of the level density, it is very easy to see that the initial condition takes the form

$$\Im z_1^{(0)}(s) = \sin \pi b_1(s) = \sin \pi(is_{12} - s_{11}) \quad (5.21)$$

and is thus according to Appendix C of the form that ensures the validity of Eq. (3.40) such that the level density is unity everywhere for all values of $\lambda$ as it should be.

For $k > 1$ the functions $b_p(s)$ have differences of eigenvalues $s_{p1}$ in the denominator which still carry imaginary increments. The exponentials of these functions have to be viewed as operators that generate distributions which complicates the integrals to be performed. However, the two level correlations can still be expressed as a double integral.

## 5.5 Two Level Correlations

For the analysis of data, the two level correlation function is of particular importance. The general considerations of Section 3.5 apply and the function $X_2(r, \lambda)$ with $r = \xi_1 - \xi_2$ is of the form (3.42). Thus, we have to evaluate $\Im z_2^{(0)}(0, t_1, t_2)$, i.e. $\Im \exp(\mp i\pi b_1(s)) \exp(\mp i\pi b_2(s))$ in the coordinates (3.41). In Appendix E we show that the initial condition takes the form

$$\Im z_2^{(0)}(0, t_1, t_2) = \frac{1}{2} \text{Re} \left( \exp\left(-i\pi \frac{t_1^2 + t_2^2}{2t_1^-}\right) - 1 \right) \quad (5.22)$$

where the exponential has to be interpreted as a power series involving the operator $1/t_1^-$. Using the principal value and the $\delta$-function, the initial condition can be written as

$$\Im z_2^{(0)}(0, t_1, t_2) = -\frac{1}{2} \sum_{n=1}^{\infty} \frac{(-1)^n (\pi(t_1^2 + t_2^2))^{2n}}{2^{2n}(2n)!(2n-1)!} \frac{\partial^{2n-1}}{\partial t_1^{2n-1}} P\left(\frac{1}{t_1}\right)$$

$$+ \frac{\pi}{2} \sum_{n=0}^{\infty} \frac{(-1)^n (\pi(t_1^2 + t_2^2))^{2n+1}}{2^{2n+1}(2n+1)!(2n)!} \frac{\partial^{2n}}{\partial t_1^{2n}} \delta(t_1) \quad (5.23)$$

which is obviously not very convenient. We therefore express the initial condition (5.22) by means of an integral transform

$$\Im z_2^{(0)}(0, t_1, t_2) = \frac{\sqrt{2\pi(t_1^2 + t_2^2)}}{4} \int_0^{\infty} I_1\left(\sqrt{2\pi(t_1^2 + t_2^2)u}\right) \frac{\exp(-\varepsilon u)\cos(t_1 u)}{\sqrt{u}} du$$

(5.24)

where $I_1(z)$ is the modified Bessel function of first order. The convergence is ensured by the infinitesimal increment $\varepsilon$. This representation allows, inserted into Eq. (3.42), the evaluation of one of the three integrals, details are given in Appendix F. We arrive at the double integral

$$X_2(r, \lambda) = \frac{4}{\pi \lambda^2} \operatorname{Im} \int_0^{\infty} du \left(\frac{1}{u + i2r/\lambda^2} + \frac{1}{u}\right) \int_0^{\infty} d\rho \, \exp\left(-\frac{\rho^2}{2\lambda^2}\right)$$

$$\frac{I_1\left(\sqrt{2\pi u}\rho\right)}{\sqrt{2\pi u}} J_2\left(\sqrt{u^2 + i2ur/\lambda^2}\rho\right) \quad (5.25)$$

in which $J_2(z)$ is the Bessel function of second order. Since the $u$ integral converges as it stands if the $\rho$ integration is done first, we have already taken the limit $\varepsilon \to 0$. Thus, for the first time, a closed formula for the two point correlation is given which is valid for all values of the transition parameter $\lambda$. Yet another double integral representation can be found

$$X_2(r, \lambda) = \frac{8r}{\pi} \operatorname{Im} \int_0^{\infty} du \, \exp(\pi \lambda^2 u) \left(u - i\frac{r}{\lambda^2}\right)$$

$$\int_0^1 d\eta \, \eta^2 \exp\left(-\left(u - i2\frac{r}{\lambda^2}\right)\frac{\lambda^2 u \eta^2}{2}\right)$$

$$\frac{J_1\left(\sqrt{2\pi}\lambda^2 u \sqrt{u - i2r/\lambda^2}\eta\right)}{\sqrt{2\pi}\lambda^2 u \sqrt{u - i2r/\lambda^2}} \quad (5.26)$$

which involves only one Bessel function. The derivation is sketched in Appendix F. According to the integral representations, the two level correlation function has the property

$$X_2(0,\lambda)\Big|_{\lambda\neq 0} = 0 \qquad (5.27)$$

which simply means that an arbitrarily small GUE admixture to the Poisson spectrum lifts immediately all degeneracies.

Our result for the two level correlation function can be checked by discussing some limit relations. First, for large energy differences, the two level correlations must approach unity,

$$\lim_{r\to\infty} X_2(r,\lambda) = 1 . \qquad (5.28)$$

Second, the limit of vanishing transition parameter must also yield unity

$$\lim_{\lambda\to 0} X_2(r,\lambda) = 1 \qquad (5.29)$$

since the Poisson spectrum is totally uncorrelated. It is shown in Appendix G that our result indeed satisfies both limit relations. Furthermore, it is important to compare our formula for small but finite values of the transition parameter to the perturbative calculations of Refs. [9, 10]. From Eqs. (3.45), (3.46) and (5.24) we find for small $\lambda$

$$X_2(r,\lambda) \simeq X_{2,0}(r/\lambda) = \frac{r}{\lambda}\int_0^\infty \exp\left(-\frac{k^2}{2}\right)\sin\frac{rk}{\lambda}\,dk \qquad (5.30)$$

in perfect agreement with the results of Refs. [9, 10]. Notice that the definitions of the transition parameter differ, $\lambda$ in Ref. [10] has to be replaced by $\lambda^2/2$ to compare the formulae. Details of the derivation of Eq. (5.30) are given in Appendix G.

Since further analytical treatment of the integrals (5.25) and (5.26) seems not possible, we resort to a numerical calculation of formula (5.25) for some values of the transition parameter $\lambda$. In Fig. 1 we show the results for the three $\lambda$ values 0.1, 0.5 and 0.7. The numerics is not completely trivial, the $\rho$ integration yields an oscillating function with large amplitudes as integrand for the $u$ integration. Therefore, for the larger $\lambda$ values 0.5 and 0.7, the calculation is only meaningful up to energy differences of $r = 1.2$.

For $\lambda = 0.1$, the uncorrelated Poisson spectrum has been affected by a rather weak level repulsion. All degeneracies are lifted, but the correlations are only relevant for smaller energy differences $r$. The function has therefore a very steep gradient near the origin $r = 0$. However, the level repulsion has to be compensated for intermediate $r$ before the Poisson situation is restored. This compensation manifests itself in in a region where $X_2(r,\lambda)$ exceeds unity which we will refer to as the overshoot. This effect has already been observed in Refs. [9, 10]. It can be understood as a consequence of a normalization or sum rule. In the Poisson case, the integral over the two level cluster function on the unfolded is zero which implies that the integral of $Y_2(r,\lambda) = 1 - X_2(r,\lambda)$ is approximately zero for small $\lambda$ thus causing the overshoot. In fact, it is easily shown that the approximation (5.30) fulfils the sum rule, i.e. the integral of $1 - X_{2,0}(r/\lambda)$ yields zero. On the other hand, the integral over the two level cluster function on the unfolded scale is unity in the GUE case. Hence, the integral of $Y_2(r,\lambda)$ changes from zero to one as the transition is made from $\lambda = 0$ to $\lambda \to \infty$. We mention in passing that on the original energy scale where the level number $N$ is finite, the two level cluster function must satisfy the same sum rule for all values of the transition parameter due to the definition (2.16). Thus, the sum rule and the limit of infinitely many levels $N \to \infty$ do not commute in the Poisson case. However, the interpretation of our results in terms of a change of the sum rule on the unfolded energy scale is very useful. It implies immediately that the overshoot for smaller $\lambda$ has to disappear as $\lambda$ is made bigger. Indeed, Fig. 1 shows that $X_2(r,\lambda)$ for $\lambda = 0.5$ exceeds unity only very slightly. For $\lambda = 0.7$, no overshoot is visible anymore.

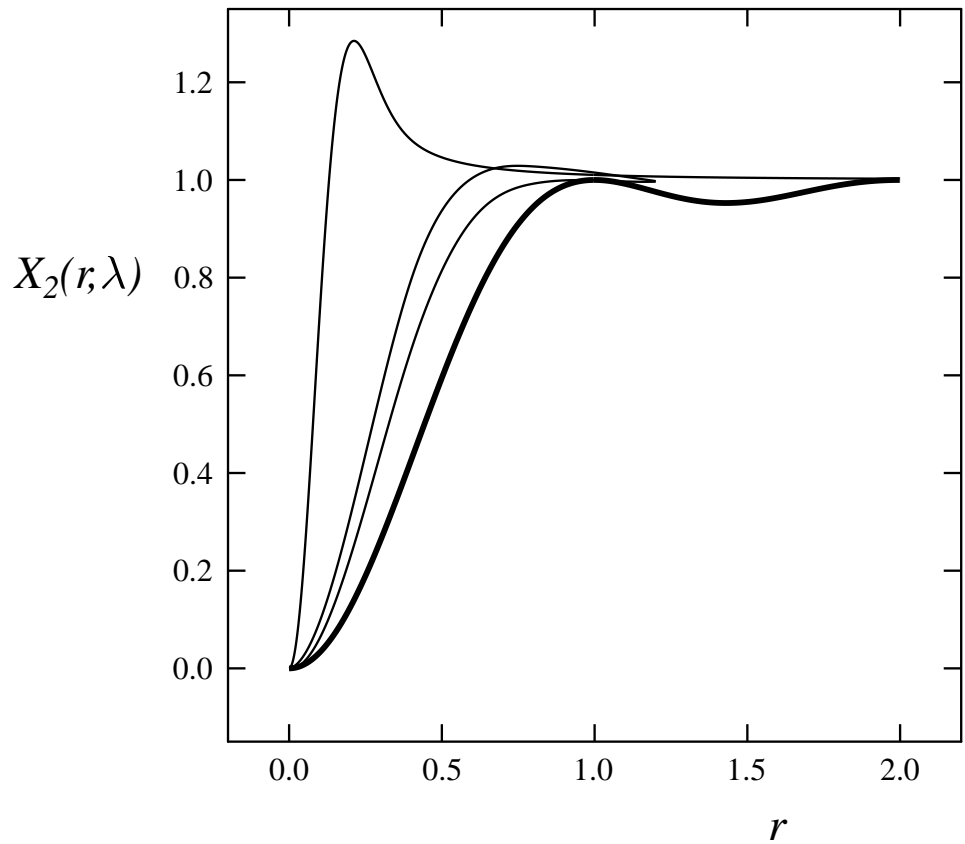

Figure 1: Two level correlation function of the transition from Poisson to GUE for different values of the transition parameter $\lambda$, calculated numerically using formula (5.25). In the Poisson case, i.e. for $\lambda = 0$, the function is unity which is not drawn. The three thin solid lines correspond from the left to the right to $\lambda$ values of 0.1, 0.5 and 0.7, respectively. The thick solid line is the pure GUE case corresponding to $\lambda \to \infty$.

Furthermore, it becomes obvious from Fig. 1 that the two level correlations reach the GUE limit rather quickly when the transition parameter $\lambda$ approaches values of about unity. This is consistent with the numerical simulations of this transition in Ref. [7]. Roughly the same speed has also been observed and analytically shown for other, but related transitions, in particular for the breaking of time reversal invariance, i.e. the GOE GUE transition [12], and the breaking of quantum numbers [15].

We mention finally that double integral representations of the level number variance $\Sigma^2(L,\lambda)$ for the transition from Poisson regularity to GUE can be found straightforwardly by using the general expression (3.48) and the methods of Appendix F. Since the results are similar to formulae (5.25) and (5.26) but more complicated, we decided not to present them here.

# 6 Summary and Discussion

We analyzed the statistical model of a system that can be written as the sum of an arbitrary and a chaotic part. This choice is motivated by a whole variety of experimental and theoretical investigations. The chaotic part was drawn from the Gaussian Ensembles whose fluctuation properties are known to describe chaos in quantum mechanics. We expressed the generating function of the spectral correlations for the transition from the arbitrary to the chaotic regime as a supersymmetric matrix-model. A Pastur equation for the level density involving the level density of the arbitrary part was found through a saddlepoint approximation. In the case of broken time reversal invariance, the graded eigenvalue method was used to derive a closed integral representation of the generating function which contains the generating function of the arbitrary correlations in the integrand. The number of integrals to be done is twice the order of the correlation. The unfolding procedure could be performed directly in this integral representation which immediately implied the translation invariance of all correlations, provided the correlations of the arbitrary part were translation invariant. Employing this insight, we showed that the level density is, as expected, unity for all values of the transition parameter. Furthermore, the two level correlation function could be written as a double integral. Consequently, a double integral representation for the level number variance was found, too.

For all three universality classes, we derived diffusion equations for the generating functions which are completely equivalent to the integral representations. The diffusion takes place in the curved space of the eigenvalues of the supermatrices. Only the radial part of the Laplacean contributes such that no anticommuting variables have to be considered explicitly. The squared transition parameter provides the diffusion time. The generating function of the arbitrary correlations is the initial condition. Hence, arbitrary correlations are propagated diffusively into the chaotic regime. The form of the diffusion equation is identical on the original and the unfolded scale which is a strong demonstration for the utility of the graded eigenvalue method. We expect that these diffusion equations are also valid if the initial condition is not purely spectral, for example, in the case of a coupling to channels. Stationary equations for the pure chaotic correlations follow directly from the diffusive process.

We discussed the transition from Poisson regularity to chaos by taking an initial condition without spectral correlations. The generating function can be worked out in closed form and was used in the general integral representation. On the unfolded energy scale, the generating function is an infinite series of distributions which could be written as an integral transform. Using the general formula, the two level correlations were expressed as a double integral. This is the first time that such an integral representation has been found. Most likely, There is no other method that allows for an evaluation of both ensemble averages and yields a low dimensional integral representation at the same time for all values of the transition parameter.

## Appendix

In Appendix A, the Pastur equation is derived. The stationary equation for the GUE is discussed in Appendix B. In Appendix C, the level density on the unfolded scale is evaluated. The two level correlations on the unfolded scale are discussed in Appendix D. In Appendix E, the Poisson initial condition for the two level correlations is worked out. Integral representations of the two level correlation function are constructed in Appendix F. In Appendix G, checks of the two level correlation function are performed.

## A  Derivation of the Pastur Equation

Since the generating function $Z_1^{(0)}(s)$ is independent of the superunitary matrix $u$, we need only the radial part of the gradient operator $\partial/\partial\sigma$. In Ref. [35] the full operator was expressed in spherical coordinates, the radial part reads

$$u^{-1}\left(\tau_{11}\frac{\partial}{\partial s_1} + i\tau_{22}\frac{\partial}{\partial s_2}\right)u \tag{A.1}$$

where the $2\times 2$ matrices $\tau_{pp}$ have unity in the entry $pp$ and zeros elsewhere. We omit the index $k=1$. From Eq. (3.21) we find that the saddlepoint $\sigma_0 = u_0^{-1} s_0 u_0$ is given by

$$0 = \mathrm{trg}\left(-\frac{2}{\alpha^2}(s_0 - x) + \left(\tau_{11}\frac{\partial}{\partial s_1} + i\tau_{22}\frac{\partial}{\partial s_2}\right)\ln Z_1^{(0)}(s)\bigg|_{s=s_0}\right)u_0 d\sigma u_0^{-1}. \tag{A.2}$$

Since the supertrace has to vanish for all matrix differentials $d\sigma$, the diagonal matrix in front of $u_0 d\sigma u_0^{-1}$ is zero and we obtain the two coupled equations

$$0 = -\frac{2}{\alpha^2}(s_{10} - x) + \frac{\partial}{\partial s_1}\ln Z_1^{(0)}(s)\bigg|_{s=s_0}$$

$$0 = -\frac{2}{\alpha^2}(is_{20} - x) + i\frac{\partial}{\partial s_2}\ln Z_1^{(0)}(s)\bigg|_{s=s_0}. \tag{A.3}$$

By setting $s_1 = x' - J'$ and $is_2 = x' + J'$ we introduce new variables $x'$ and $J'$ which play formally the role of energy and source variable in the generating function $Z_1^{(0)}(s)$. The difference of the Eqs. (A.3) yields

$$0 = -\frac{2}{\alpha^2}(s_{10} - is_{20}) + \frac{\partial}{\partial x'}\ln Z_1^{(0)}(s)\bigg|_{s=s_0} \tag{A.4}$$

which is solved by $s_{10} = is_{20}$, i.e. $J' = 0$, because the derivative of the generating function with respect to the energy is zero for vanishing source term. The sum of the Eqs. (A.3),

$$0 = +\frac{2}{\alpha^2}(s_{10} + is_{20} - 2x) + \frac{\partial}{\partial J'}\ln Z_1^{(0)}(s)\bigg|_{s=s_0}, \tag{A.5}$$

involves now the derivative of the generating function with respect to the source variable at $J' = 0$ which is according to Eq. (2.6) nothing else but the level density. This gives Eq. (3.22).

In order to integrate over the massive modes we have to expand up to second order,

$$L(\sigma_0 + d\sigma, x + J, \alpha) = L(\sigma_0, x + J, \alpha) + \frac{1}{2}\mathrm{trg}\frac{\partial^2 L}{\partial \sigma^2}\bigg|_{\sigma=\sigma_0} d\sigma^2, \tag{A.6}$$

and evaluate the Gaussian integral over $d\sigma$. It is straightforward to work out the second order term explicitly. However, this is not necessary since the gradient is Hermitean and the Lagrange function is well behaved. Thus, the matrix $\partial L/\partial \sigma^2$ at $\sigma = \sigma_0$ is also Hermitean and can be absorbed into $d\sigma$. Consequently, the Gaussian integral converges and yields just unity. The generating function

$$\begin{aligned} Z_1(x+J,\alpha) &= \exp L(\sigma_0, x+J, \alpha) \\ &= \exp\left(-\frac{1}{\alpha^2}\operatorname{trg}(s_0 - x - J)^2\right) \qquad (N \to \infty) \end{aligned} \qquad (A.7)$$

gives the result (3.23).

## B  Stationary Equation for the GUE

The ansatz (4.9) reduces the diffusion (4.8) in a curved space to a diffusion in a Cartesean space. Similarly, the ansatz $Z^{(1)}_{2k}(r) = W^{(1)}_{2k}(r)/B_k(r)$ reduces Eq. (4.33) to the Cartesean stationary equation

$$\left(\frac{1}{2}\frac{\partial^2}{\partial \vec{r}^2} + \vec{r}\cdot\frac{\partial}{\partial \vec{r}} + k\right) W^{(1)}_{2k}(r) = 0 \qquad (B.1)$$

where we used that $\vec{r}\cdot\partial \ln B_k(r)/\partial \vec{r}$ is equal to the constant $-k$. An equivalent form of this equation is given by

$$\left(\frac{\partial}{\partial \vec{r}} + \vec{r}\right)^2 W^{(1)}_{2k}(r) = \vec{r}^2 W^{(1)}_{2k}(r) \qquad (B.2)$$

whose left hand side suggests the ansatz $W^{(1)}_{2k}(r) = \exp(-\vec{r}^2/2)V^{(1)}_{2k}(r)$. It leads to the equation

$$\left(\frac{\partial^2}{\partial \vec{r}^2} - \vec{r}^2\right) V^{(1)}_{2k}(r) = 0 \qquad (B.3)$$

which is of oscillator type. Obviously, the eigenvalues $r_{p1}$ and $ir_{q2}$ can be arbitrarily paired such that the solution can be built with functions $v^{(1)}(r_{p1}, ir_{q2})$ that satisfy

$$\left(\frac{\partial^2}{\partial r_{p1}^2} - r_{p1}^2\right) v^{(1)}(r_{p1}, ir_{q2}) = \left(\frac{\partial^2}{\partial (ir_{q2})^2} - (ir_{q2})^2\right) v^{(1)}(r_{p1}, ir_{q2}) . \qquad (B.4)$$

The equation is separable and the solution is a product. The addition of the term $(2n+1)v^{(1)}(r_{p1}, ir_{q2})$ to the equation makes both sides to the oscillator wave equation of index $n$. The imaginary unit in front of $ir_{q2}$ takes care of a proper cancelation of the $n$-dependent parts. Thus we find for arbitrary $n$ the solution

$$v^{(1)}(r_{p1}, ir_{q2}) = \widehat{\varphi}_n(r_{p1})\widehat{\varphi}_n(ir_{q2}) \qquad (B.5)$$

where the functions $\widehat{\varphi}_n$ are a linear combination of both fundamental solutions of the oscillator equation [25]. Every linear combination of these product solutions is again a solution, in particular we can choose

$$v^{(1)}(r_{p1}, ir_{q2}) = \widehat{K}_N(r_{p1}, ir_{q2}) = \sum_{n=0}^{N-1} \widehat{\varphi}_n(r_{p1}) \varphi_n(ir_{q2}) \qquad (B.6)$$

for arbitrary $N$. Moreover, all boundary conditions are satisfied by

$$V_{2k}^{(1)}(r) = \det[\widehat{K}_N(r_{p1}, ir_{q2})]_{p,q=1,\ldots,k} \qquad (B.7)$$

which is indeed the correct result. Since none of the equations involved is explicitly $N$-dependent, one can also construct envelope solutions corresponding to an asymptotic expansion in $N$.

## C  Level Density on the Unfolded Scale

By construction, the level density of the arbitrary ensemble has to be unity on the unfolded scale, $X_1^{(0)}(\xi) = 1$. According to Eq. (2.6), this is only possible if the generating function has a power series expansion of the form

$$\Im z_1^{(0)}(s) = -\pi(s_1 - is_2) + \sum_{n=2}^{\infty} c_n (s_1 - is_2)^n \qquad (C.1)$$

with some coefficients $c_n$. We drop the index $k = 1$ in our notation. For the level density of the transition, we find from Eq. (3.35) the double integral

$$X_1(\xi, \lambda) = -\frac{1}{(\pi\lambda)^2} \int_{-\infty}^{+\infty}\int_{-\infty}^{+\infty} \exp\left(-\frac{1}{\lambda^2}\left((s_1-\xi)^2 - (is_2-\xi)^2\right)\right) \Im z_1^{(0)}(s) \frac{ds_1 ds_2}{s_1 - is_2}$$

$$= \frac{1}{\pi\lambda^2} \int_{-\infty}^{+\infty}\int_{-\infty}^{+\infty} \exp\left(-\frac{1}{\lambda^2}\left(s_1^2 + s_2^2\right)\right)$$

$$\left(1 - \frac{1}{\pi}\sum_{n=2}^{\infty} c_n(s_1 - is_2)^{n-1}\right) ds_1 ds_2 . \qquad (C.2)$$

Apart from the Gaussian functions, the integrand depends only on the difference $s_1 - is_2$, the sum $s_1 + is_2$ does not appear. Thus, we can apply the formula

$$\frac{1}{\pi\lambda^2} \int_{-\infty}^{+\infty}\int_{-\infty}^{+\infty} \exp\left(-\frac{1}{\lambda^2}(s_1^2 + s_2^2)\right) f(s_1 - is_2) ds_1 ds_2 = f(0) \qquad (C.3)$$

which holds for any well behaved function $f$. This is a well known mean value theorem from complex analysis. It can be shown by introducing polar coordinates $s_1 + is_2 = u \exp(i\vartheta)$ and then expanding the integrand in the phase. In the limit $\lambda \to 0$, it is also valid due to the occurrence of $\delta(s_1)\delta(s_2)$. Formula (C.3) yields immediately the result (3.40).

## D  Two Level Correlations on the Unfolded Scale

With the new integration variables (3.41), the relevant parts of the integrand read

$$\text{trg}(s-\xi)^2 = \frac{1}{2}(s_1^2 + s_2^2 + t_1^2 + t_2^2) - \widetilde{r}(s_1 - is_2) - r(t_1 - it_2)$$

$$B_2(s) = \frac{4t_1}{(s_1 - is_2 + t_1 - it_2)(s_1 - is_2 + t_1 + it_2)}$$

$$\frac{4it_2}{(s_1 - is_2 - t_1 - it_2)(s_1 - is_2 - t_1 + it_2)} \quad . \text{(D.1)}$$

As in the case of the level density in Appendix C, apart from the Gaussian functions, the whole integrand depends only on the difference $s_1 - is_2$. We can thus use formula (C.3) and find

$$X_2(r,\lambda) = \frac{8}{\pi^3 \lambda^2} \int_{-\infty}^{+\infty}\int_{-\infty}^{+\infty} \exp\left(-\frac{1}{2\lambda^2}(t_1^2 + t_2^2) - \frac{r}{\lambda^2}(t_1 - it_2)\right)$$

$$\frac{t_1 it_2}{(t_1^2 + t_2^2)^2} \Im z_2^{(0)}(0, t_1, t_2)\, dt_1 dt_2 \qquad \text{(D.2)}$$

which does not contain $\widetilde{r}$ anymore as required. Since the integrand is odd in each of the variables $t_1$ and $t_2$ except of the $r$ dependent exponentials, it is enough to keep the odd parts of those which gives formula (3.42).

## E  Poisson Initial Condition for the Two Level Correlations

In the case $k = 2$ we have to consider the functions

$$b_1(s) = \frac{(is_{12} - s_{11})(is_{22} - s_{11})}{s_{21}^{\pm} - s_{11}^{\pm}} \quad \text{and} \quad b_2(s) = \frac{(is_{12} - s_{21})(is_{22} - s_{21})}{s_{11}^{\pm} - s_{21}^{\pm}}$$
(E.1)

in the exponent. We introduce the short hand notation $b_p^{\pm\pm}$ indicating the sign of the imaginary increments. From Eq. (2.5) we find the expression

$$(i2)^2 \Im z_2^{(0)}(s) = \exp(+i\pi(b_1^{--} + b_2^{--})) + \exp(-i\pi(b_1^{++} + b_2^{++}))$$

$$- \exp(-i\pi(b_1^{+-} - b_2^{+-})) - \exp(+i\pi(b_1^{-+} - b_2^{-+})) \,. \text{ (E.2)}$$

which can be simplified with the sum rule (5.12) giving

$$\Im z_2^{(0)}(s) = \frac{1}{2}\text{Re}\left(\exp(i\pi(b_1^{-+} - b_2^{-+})) - \exp(-i\pi\text{tr}gs)\right)$$

$$= \frac{1}{2}\text{Re}\exp(-i\pi\text{tr}gs)\left(\exp(-i2\pi b_2^{-+}) - 1\right) . \quad \text{(E.3)}$$

We use the coordinates (3.41) and put $s_1 - is_2 = 0$. The supertrace vanishes due to $\text{tr}gs = s_1 - is_2$ and we arrive at Eq. (5.22). Obviously, we have to deal with a function of the operator $1/t_1^-$ which is defined through its power series expansion. By expanding the exponential and using

$$\frac{1}{(t_1^-)^n} = \frac{(-1)^{n-1}}{(n-1)!}\frac{\partial^{n-1}}{\partial t_1^{n-1}}\frac{1}{t_1^-} = \frac{(-1)^{n-1}}{(n-1)!}\frac{\partial^{n-1}}{\partial t_1^{n-1}}\left(P\left(\frac{1}{t_1}\right) + i\pi\delta(t_1)\right) \quad \text{(E.4)}$$

we arrive at Eq. (5.23).

## F   Integral Representations of the Two Level Correlations

The integral (5.24) is a special case of formula 6.643.2 given in Ref. [37]. It can also be derived directly by expanding the exponential in Eq. (5.22), using Eq. (E.4) and the formula

$$\frac{1}{t_1^-} = i\int_0^\infty \exp(-it_1^- u)\,du . \quad \text{(F.1)}$$

Equation (5.24) is inserted into the integral representation (3.42) and polar coordinates $t_1 = \rho\cos\vartheta$, $t_2 = \rho\sin\vartheta$ are introduced. The $\vartheta$ integration is of the form

$$\int_0^{2\pi}\exp(ia\cos\vartheta + ib\sin\vartheta)\cos\vartheta\sin\vartheta\,d\vartheta = -2\pi\frac{\partial^2}{\partial a\partial b}J_0\left(\sqrt{a^2+b^2}\right)$$

$$= 2\pi\frac{ab}{a^2+b^2}J_2\left(\sqrt{a^2+b^2}\right) \quad \text{(F.2)}$$

where $J_n(z)$ is the Bessel function of order $n$. In our case we have $a = -\rho(u - ir/\lambda^2)$ and $b = \rho r/\lambda^2$. Collecting everything yields the integral representation (5.25). The $\rho$ integration which has the form

$$M(a,b,c) = \int_0^\infty \exp(-c\rho^2)\,I_1(a\rho)\,J_2(b\rho)\,d\rho \quad \text{(F.3)}$$

can be rewritten by making use of Sonine's formula [38]

$$J_{\mu+\nu+1}(z) = \frac{z^{\nu+1}}{2^\nu\Gamma(\nu+1)}\int_0^{\pi/2}J_\mu(z\sin\vartheta)\,\sin^{\mu+1}\vartheta\cos^{2\nu+1}\vartheta\,d\vartheta \quad \text{(F.4)}$$

for $\mu = 1$ and $\nu = 0$,

$$J_2(z) = z \int_0^1 J_1(z\eta) \eta^2 \, d\eta \tag{F.5}$$

where we changed the integration variable to $\eta = \sin\vartheta$. The order of the Bessel function $J_2(b\rho)$ is thus reduced by one which allows the application of Weber's formula [38]

$$\int_0^\infty \exp(-c\rho^2) \, I_\nu(a\rho) \, J_\nu(b\rho) \, \rho \, d\rho = \frac{1}{2c} \exp\left(\frac{a^2 - b^2}{4c}\right) J_\nu\left(\frac{ab}{2c}\right) \tag{F.6}$$

for $\nu = 1$. Collecting everything we find

$$M(a,b,c) = \frac{b}{2c} \exp\left(\frac{a^2}{4c}\right) \int_0^1 \exp\left(-\frac{b^2\eta^2}{4c}\right) J_1\left(\frac{ab}{2c}\eta\right) \eta^2 \, d\eta \tag{F.7}$$

which yields with $a = \sqrt{2\pi u}$, $b = \sqrt{u^2 - i2ur/\lambda^2}$ and $c = 1/2\lambda^2$ the integral representation (5.26).

## G  Checks of the Two Level Correlations

We start with a general expression for the two level correlation function, but instead of Eq. (3.42) we use the equivalent form (D.2) and shift the integration variables according to $t_1 \to t_1 + r$ and $t_2 \to t_2 - ir$ which gives

$$X_2(r,\lambda) = \frac{i8}{\pi^3 \lambda^2} \int_{-\infty}^{+\infty}\int_{-\infty}^{+\infty} \exp\left(-\frac{1}{2\lambda^2}(t_1^2 + t_2^2)\right) \frac{(t_1 - r)(t_2 + ir)}{(t_1 - it_2)^2((t_1 - r) - i(t_2 + ir))^2}$$
$$\Im z_2^{(0)}(0, t_1 - r, t_2 + ir) \, dt_1 dt_2 \, . \tag{G.1}$$

The contribution from the Berezinian yields $i/4(t_1 - it_2)^2$ in the limit of large energy differences $r$. For a Poisson initial condition, we have from Eq. (5.22)

$$\Im z_2^{(0)}(0, t_1 - r, t_2 + ir) = \frac{1}{2}\text{Re}\left(\exp\left(-i\frac{\pi}{2}(t_1 - it_2)\left(1 + i\frac{t_2 + ir}{t_1 - r}\right)\right) - 1\right) \tag{G.2}$$

where now, for consistency reasons, the imaginary unit in front of $t_2$ which appeared due to the shift has to be ignored when taking the real part. Hence, the limit of large $r$ yields

$$\frac{1}{2}\text{Re}\left(\exp\left(-i(t_1 - it_2)\right) - 1\right) = -\sin^2\frac{\pi}{2}(t_1 - it_2) \, . \tag{G.3}$$

Collecting everything we find

$$\lim_{r \to \infty} X_2(r, \lambda) = \frac{1}{\pi^3 \lambda^2} \int\limits_{-\infty}^{+\infty}\int\limits_{-\infty}^{+\infty} \exp\left(-\frac{1}{2\lambda^2}(t_1^2 + t_2^2)\right) \frac{\sin^2(\pi(t_1 - it_2)/2)}{(t_1 - it_2)^2} dt_1 dt_2 \tag{G.4}$$

which gives the limit relation (5.28) by applying the integral formula (C.3).

The representation (G.1) of the two level correlation function allows to consider the limit of vanishing transition parameter, too. For $\lambda \to 0$ the Gaussians become the functions $\delta(t_1)\delta(t_2)$. After integration we find

$$\lim_{\lambda \to 0} X_2(r, \lambda) = \frac{4}{\pi^2} \lim_{t_1, t_2 \to 0} \frac{\sin^2(\pi(t_1 - it_2)/2)}{(t_1 - it_2)^2} = 1 \tag{G.5}$$

and thus the limit relation (5.29).

Finally, we discuss the two level correlation function for small but finite values of the transition parameter. From Eq. (5.24) we have to zeroth order

$$\left.\frac{\partial \Im z_2^{(0)}(0, \lambda t_1, \lambda t_2)}{\partial(\lambda^2)}\right|_{\lambda^2 = 0} = \frac{\pi^2(t_1^2 + t_2^2)^2}{16} \frac{\partial}{\partial t_1} P\left(\frac{1}{t_1}\right) \tag{G.6}$$

which has to be inserted into formula (3.46) for $n = 0$. Luckily, there are some cancelation that decouple the two integrations to be performed. We arrive at

$$\begin{aligned} X_{2,0}(r/\lambda) &= M_1(r/\lambda) M_2(r/\lambda) \\ M_1(\eta) &= \frac{1}{\sqrt{2\pi}} \int\limits_{-\infty}^{+\infty} \exp\left(-\frac{t_1^2}{2}\right) t_1 \sinh \eta t_1 \frac{\partial}{\partial t_1} P\left(\frac{1}{t_1}\right) dt_1 \\ M_2(\eta) &= \frac{1}{\sqrt{2\pi}} \int\limits_{-\infty}^{+\infty} \exp\left(-\frac{t_2^2}{2}\right) t_2 \sin \eta t_2 \, dt_2 \; . \end{aligned} \tag{G.7}$$

Integration by parts yields three ordinary principal value integrals for $M_1(\eta)$, two of them cancel each other and we find

$$M_1(\eta) = -\frac{1}{\sqrt{2\pi}} \int\limits_{-\infty}^{+\infty} \exp\left(-\frac{t_1^2}{2}\right) \frac{\sinh \eta t_1}{t_1} dt_1 = -\int\limits_0^\eta \exp\left(\frac{\eta'^2}{2}\right) d\eta' , \tag{G.8}$$

while the second integral gives trivially

$$M_2(\eta) = \eta \exp\left(-\frac{\eta^2}{2}\right) . \tag{G.9}$$

Thus, the product contains Dawson's integral [37],

$$X_{2,0}(r/\lambda) = \frac{r}{\lambda} \exp\left(-\frac{r^2}{2\lambda^2}\right) \int_0^{r/\lambda} \exp\left(\frac{\eta^2}{2}\right) d\eta \qquad (G.10)$$

which is identical to the result (5.30).

## Acknowledgments

It is a pleasure for me to thank S. Creagh, S. Tomsovic and H.A. Weidenmüller for stimulating discussions. I am especially grateful to F. Leyvraz for his crucial help in constructing the expansion of the two level correlation function in terms of the transition parameter. I acknowledge financial support from the Max Planck Gesellschaft as an Otto Hahn fellow and from the Danish Research Council.